\documentclass[apj,twocolappendix,twocolumn]{openjournal}

\usepackage{newtxtext,newtxmath,enumitem,multirow,ctable,CJK}
\usepackage[T1]{fontenc}
\usepackage[breaklinks,color links,citecolor=blue,urlcolor=blue]{hyperref}

\newcommand{\sect}{Section~}

\newcommand{\figu}{Fig.~}
\newcommand{\figus}{Figs.~}
\newcommand{\tab}{Table~}

\newcommand{\eq}{Eq.~}

\newcommand{\athenak}{\texttt{AthenaK}}
\newcommand{\athenapp}{\texttt{Athena++}}
\newcommand{\kokkos}{\texttt{Kokkos}}

\newcommand{\orcidauthor}[3]{\author{\href{http://orcid.org/#1}{#2 \openin1 Orcid-ID.png \ifeof1 \else \hskip2pt\includegraphics[width=9pt]{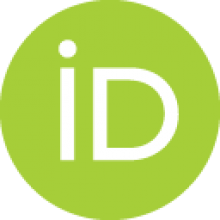}\fi}$^{#3}$}}
\shorttitle{Accretion Disk with Toroidal Fields}
\shortauthors{Guo et al.}
\graphicspath{{./}{figures/}}

\begin{document}

\title{Idealized Global Models of Accretion Disks with Strong Toroidal Magnetic Fields}

\begin{CJK*}{UTF8}{gbsn}

\email{* mhguo@princeton.edu}

\orcidauthor{0000-0002-3680-5420}{Minghao Guo (郭明浩)}{1,*}
\orcidauthor{0000-0001-9185-5044}{Eliot Quataert}{1}
\orcidauthor{0000-0001-8479-962X}{Jonathan Squire}{2}
\orcidauthor{0000-0003-3729-1684}{Philip F. Hopkins}{3}
\orcidauthor{0000-0001-5603-1832}{James M. Stone}{1,4}

\affiliation{$^{1}$Department of Astrophysical Sciences, Princeton University, Princeton, NJ 08544, USA}
\affiliation{$^{2}$Physics Department, University of Otago, 730 Cumberland St., Dunedin 9016, New Zealand}
\affiliation{$^{3}$TAPIR, Mailcode 350-17, California Institute of Technology, Pasadena, CA 91125, USA}
\affiliation{$^{4}$School of Natural Sciences, Institute for Advanced Study, 1 Einstein Drive, Princeton, NJ 08540, USA}

\begin{abstract}
We present global magnetohydrodynamic (MHD) simulations of idealized accretion disks with a strong toroidal magnetic field using an equation of state that fixes the gas thermal scale height. The disk forms from the inflow of a rotating magnetized gas cloud with a toroidal magnetic field.
We find that the system maintains a moderately strong mean azimuthal field in the midplane, with plasma-$\beta\sim1$, trans-Alfv\'enic fluctuations, and large accretion stresses $\alpha\sim0.1$.
The azimuthal field in the disk is continuously escaping along the vertical direction but is also replenished via a local dynamo.
The inflowing gas initially forms a strongly magnetized Keplerian disk with $\beta\ll1$ and $\alpha \gg 1$. The disk gradually collapses from the inside out over $\sim 50-80$ orbits to form a moderately magnetized disk with $\beta\sim1$ and $\alpha\sim0.1$. Radial advection of azimuthal magnetic field can maintain $\beta\lesssim1$ exterior to the circularization radius but not inside of it. Inclusion of a net initial vertical magnetic field can lead to an even more strongly magnetized disk midplane, consistent with previous work. When the gas thermal scale is not resolved ($\lesssim 4$ cells per thermal scale height), however, the disk remains highly magnetized with $\beta \ll 1 $.
We discuss our results in the context of related shearing box simulations and other global disk simulations. The level of angular momentum transport found here is consistent with that inferred observationally in dwarf novae and X-ray transient outbursts, unlike simulations of weakly magnetized accretion disks.
\end{abstract}

\keywords{Accretion (14) --- Black holes (162) --- Supermassive black holes (1663) --- Active galactic nuclei (16) --- Quasars (1319) --- X-ray binary stars (1811) --- Astrophysical fluid dynamics (101) --- Magnetohydrodynamics (1964) --- Magnetohydrodynamical simulations (1966)}

\section{Introduction} \label{sec:intro}

The classic analytic model of accretion disks is the ``$\alpha$-disk'' model proposed in the seminal work by \citet{Shakura1973A&A....24..337S}, where a parameter $\alpha$ quantifies the effective viscosity and thus the rate of angular momentum transport.
For hotter systems in which the accreting plasma is ionized, angular momentum transport is believed to be dominated by magnetic stresses produced by either winds \citep{Blandford1982MNRAS.199..883B} or local turbulence induced by the magnetorotational instability \citep[MRI; ][]{Balbus1991ApJ...376..214B}. In local and global numerical simulations, the angular momentum transport produced by the MRI corresponds to $\alpha$ between $\simeq 0.001$ and $\simeq 1$, depending on the physical parameters and resolution, with net vertical magnetic flux leading to the larger values of $\alpha$ ~\citep{Hawley1995ApJ...440..742H, Stone1996ApJ...463..656S, Bai2013ApJ...767...30B, Ryan2017ApJ...840....6R}. Transport inferred observationally in the outbursts of dwarf novae and X-ray binaries favors $\alpha \sim 0.1-0.3$~\citep{King2007MNRAS.376.1740K, Tetarenko2018Natur.554...69T}, corresponding to a strongly magnetized accretion disk. In addition, magnetic pressure support can stabilize the disk against thermal and viscous instabilities~\citep{Frank2002apa..book.....F, Begelman2007MNRAS.375.1070B}, which would otherwise be generic in the inner parts of luminous black hole and neutron star accretion disks (but there is little direct phenomenological evidence for such instabilities). Strong magnetic pressure support also produces lower gas densities for a given accretion rate, making the disks in active galactic nuclei (AGN) less susceptible to gravitational instability and star formation~\citep{Shlosman1987Natur.329..810S, Goodman2003, Johansen2008AA...490..501J}.

The most widely studied models of strongly magnetized accretion disks are those with net vertical magnetic fields. In local shearing box magnetohydrodynamic (MHD) simulations with an initial net vertical flux (NVF) with plasma-$\beta \lesssim 10^3$ (the ratio of thermal to magnetic pressure), the MRI saturates with stronger magnetic fields and more efficient angular momentum transport; for initial $\beta \lesssim 100$ the midplane of the disk becomes magnetically dominated~\citep{Bai2013ApJ...767...30B, Salvesen2016MNRAS.460.3488S.poloidal}.
Similar results are also found in global simulations~\citep{Mishra2020MNRAS.492.1855M, Kudoh2020ApJ...904....9K} where significant angular momentum transport by MHD winds can play an important role as well~\citep{Zhu2018ApJ...857...34Z} (see \citealt{Lesur2013A&A...550A..61L} for the role of winds in the shearing box).

An alternative to disks with strong vertical magnetic fields is those with strong toroidal magnetic fields. Indeed, a series of global numerical simulations have found accretion disks dominated by strong, superthermal toroidal magnetic fields with $\beta \ll 1$ and $\alpha \gtrsim 1$ due to the accompanying magnetic stress \citep{Machida2006PASJ...58..193M, Gaburov2012ApJ...758..103G, Sadowski2016MNRAS.459.4397S, Hopkins2024OJAp....7E..18H, Hopkins2024OJAp....7E..19H, Guo2024arXiv240511711G, Hopkins2025OJAp....8E..48H, Wang2025arXiv250403874W, Jiang2025arXiv250509671J}.
Theoretically, such disks may form robustly if the disk forms via compression along the rotation axis of the system (this conserves $B_z$ but amplifies $B_\phi$). However, whether or not such a configuration can in fact be maintained in the face of buoyant loss of magnetic field~\citep{Parker1958ApJ...128..664P} has been the subject of debate.
\citet{Johansen2008AA...490..501J} found that zero-net-vertical-flux (ZNVF) shearing boxes could maintain a $\beta\sim1$ toroidal field mediated by a Parker instability-driven dynamo; however, their vertical boundary conditions precluded azimuthal magnetic flux loss and  \citet{Salvesen2016MNRAS.460.3488S.poloidal} and \citet{Fragile2017MNRAS.467.1838F} showed numerical examples where strong toroidal magnetic fields cannot be sustained in the presence of Parker-instability mediated loss of magnetic flux.

\citet{Squire2024rapidstronglymagnetizedaccretion} recently reinvestigated the strongly magnetized disk in ZNVF shearing box simulations with toroidal magnetic fields. They found two regimes of accretion: (1) a weakly magnetized state (the "high $\beta$ state") with $\beta \sim 100$ and $\alpha \sim 0.01$ (as in previous work) and (2) a regime of strongly magnetized accretion (the ``low $\beta$ state'') characterized by a $\beta\sim1$ midplane dominated by a coherent azimuthal field with strong turbulence and large accretion stress $\alpha\sim1$. In the latter regime, the loss of magnetic flux due to magnetic buoyancy is compensated for by a local dynamo, maintaining the $\beta\sim1$ state. This shearing box result bears some resemblance to the global results \citep{Gaburov2012ApJ...758..103G, Hopkins2024OJAp....7E..18H, Guo2024arXiv240511711G} except that in the latter cases the magnetic field is much stronger with a plasma-$\beta\ll1$. The origin of this important quantitative difference is unclear.

To understand what conditions are required to support a superthermal magnetic field, we undertake idealized 3D global MHD simulations of accretion disks with strong toroidal magnetic fields. 
The central object is represented by a fixed point-mass gravitational potential and general relativistic effects are not included. The initial condition is not an equilibrium disk; instead, we begin with a rotating magnetized gas cloud with nearly uniform density and constant angular momentum at large radii. The gas falls inward, circularizes because of its finite angular momentum, and forms a disk whose magnetic field is initially dominated by the toroidal component.
The global geometry allows us to follow gas inflow and circularization, radial transport of magnetic field, and vertical stratification in a single calculation, while retaining enough resolution for comparison with stratified shearing-box simulations.
We intentionally choose a simple model problem in which the thermodynamics of the gas is specified by a locally (at a given radius) isothermal equation of state corresponding to a fixed disk aspect ratio, i.e., a fixed scale height relative to the local disk radius.
As we shall show, we find a regime with $\beta\sim1$ and $\alpha\sim 0.1$, similar in many respects to the low $\beta$ state in \citet{Squire2024rapidstronglymagnetizedaccretion}. Because of the uncertainty in the literature about the existence and properties of $\beta\lesssim1$ toroidally dominated disk without net vertical flux, we intentionally provide a detailed, and we hope, comprehensive analysis of the disk properties we find here and their dependence on resolution.

The rest of this article is organized as follows. In \sect\ref{sec:method} we describe the physical setup, initial and boundary conditions, and numerical method. \sect\ref{sec:results} presents the formation and magnetic transition of the disk, the resulting disk structure and angular momentum transport, the mechanism that sustains the toroidal field, and a set of robustness tests. In \sect\ref{sec:discussion}, we summarize our findings, discuss their implications, and compare them with related local and global simulations.

\begin{deluxetable}{lcccccc}
    \tablenum{1}
    \tablecaption{List of runs and their parameters. \label{tab:runs}}
    \tablewidth{0pt}
    \tablehead{
    \colhead{Model} & \colhead{$h$} & \colhead{$\beta_\phi$} & \colhead{$\beta_z$} & \colhead{Cells/} & \colhead{$hr/$} & \colhead{Runtime} 
    \\ \colhead{Name} & & & & \colhead{Level} & $\Delta x$ & \colhead{$[10^3\,t_0]$}
    }
    \startdata
    \textbf{h05n256 (fid.)} & $0.05$ & $1$ & $\infty$ & $256^3$ & $6.4$ & $0$-$60$
    \\h05n512t3\tablenotemark{a} & $0.05$ & $1$ & $\infty$ & $512^3$ & $12.8$ & $30$-$33$
    \\h05n128 & $0.05$ & $1$ & $\infty$ & $128^3$ & $3.2$ & $0$-$60$
    \\h05n64 & $0.05$ & $1$ & $\infty$ & $64^3$ & $1.6$ & $0$-$60$
    \\h10n256 & $0.1$ & $1$ & $\infty$ & $256^3$ & $12.8$ & $0$-$15$
    \\h10n128 & $0.1$ & $1$ & $\infty$ & $128^3$ & $6.4$ & $0$-$30$
    \\h10n64 & $0.1$ & $1$ & $\infty$ & $64^3$ & $3.2$ & $0$-$30$
    \\h02n256 & $0.025$ & $1$ & $\infty$ & $256^3$ & $3.2$ & $0$-$15$
    \\h02n128 & $0.025$ & $1$ & $\infty$ & $128^3$ & $1.6$ & $0$-$30$
    \\h02n64 & $0.025$ & $1$ & $\infty$ & $64^3$  & $0.8$ & $0$-$30$
    \\h01n256 & $0.0125$ & $1$ & $\infty$ & $256^3$ & $1.6$ & $0$-$15$
    \\h01n128 & $0.0125$ & $1$ & $\infty$ & $128^3$ & $0.8$ & $0$-$30$
    \\h01n64 & $0.0125$ & $1$ & $\infty$ & $64^3$  & $0.4$ & $0$-$30$
    \\ \hline
    h05b10 & $0.05$ & $0.1$ & $\infty$ & $256^3$ & $6.4$ & $0$-$30$
    \\h05b1z1 & $0.05$ & $1$ & $2$ & $256^3$ & $6.4$ & $0$-$30$
    \\h05b0z1 & $0.05$ & $\infty$ & $2$ & $256^3$ & $6.4$ & $0$-$30$ 
    \\h05ppm & $0.05$ & $1$ & $\infty$ & $256^3$ & $6.4$ & $0$-$10$
    \\h05n64-256 & $0.05$ & $1$ & $\infty$ & $64^3$ & $1.6$ & $15$-$20$
    \\h05n64-256-64 & $0.05$ & $1$ & $\infty$ & $64^3$ & $1.6$ & $20$-$30$
    \\h05n512t0\tablenotemark{b} & $0.05$ & $1$ & $\infty$ & $512^3$ & $12.8$ & $4$-$7$
    \\h10n128turb & $0.1$ & $1$ & $\infty$ & $128^3$ & $6.4$ & $0$-$30$
    \enddata
    \tablecomments{Main parameters of simulations in this work, listed approximately in the order in which they are discussed. Here $h$ is the parameter controlling the thermal scale height (see \eq\ref{eq:ic_temp}), $\beta_\phi\equiv \int P_\mathrm{gas}dV / \int B_\phi^2/2dV$ and $\beta_z\equiv \int P_\mathrm{gas}dV / \int B_z^2/2dV$. Resolution (``Cells/Level'') is specified as $N_x\times N_y\times N_z$. All simulations use 9 levels of mesh refinement (see \sect\ref{subsec:numerical_method}). The model h05n256 is the fiducial run. The value of $hr/\Delta x$ indicates a typical resolution in terms of the number of cells per thermal scale height. The actual resolution varies by a factor of 2 across the boundary of mesh refinement and is shown in \figu\ref{fig:resolution} in Appendix \ref{app:resolution}. The model h05ppm uses the piecewise parabolic (PPM) reconstruction method.}
    \tablenotetext{a}{Restarts from model h05n256 at $t=30,000\,t_0$ and runs to $t=33,000\,t_0$ (see \sect\ref{subsec:runs}).}
    \tablenotetext{b}{Restarts from model h05n256 at $t=4000\,t_0$ and runs to $t=7000\,t_0$ (see \sect\ref{subsec:miscellaneous}).}
\end{deluxetable}

\begin{figure*}[ht]
\centering
\includegraphics[width=\linewidth]{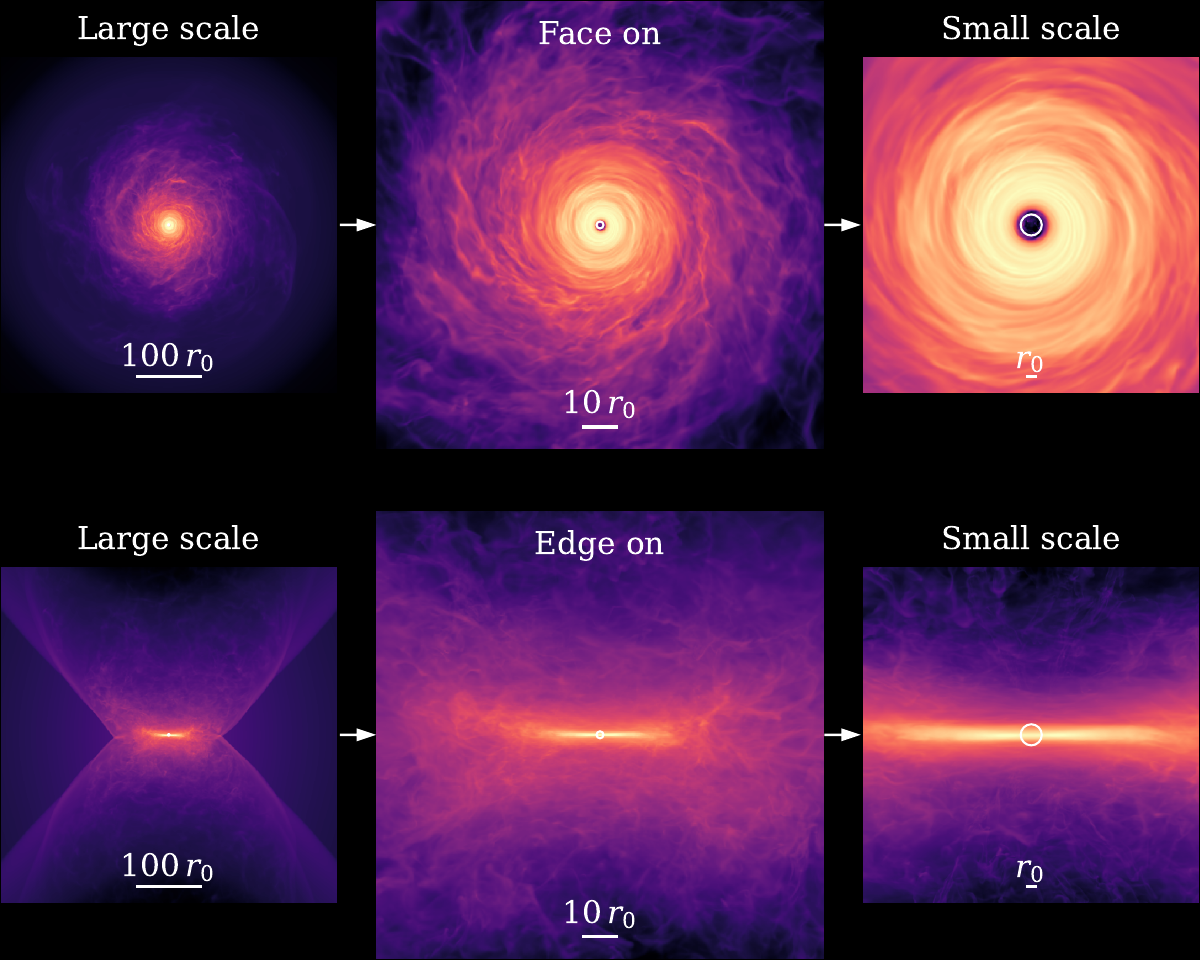}
\caption{Face-on (top panels, view from positive $z$ axis) and edge-on (bottom panels, view from positive $x$ axis) view of surface density on a logarithmic scale from large (left panels) to small (right panels) spatial scales ($h=0.05$, model h05n512t3 at $t/t_0=3.3\times10^4$). The material from large scales ($\gtrsim 100\,r_0$) inflows radially and circularizes at $R_\mathrm{circ}=16\,r_0$, forming a rotationally supported turbulent accretion disk. The white circles mark the unit radius $r_0$.
\label{fig:proj}}

\end{figure*}
\begin{figure*}[ht]
\centering
\includegraphics[width=\linewidth]{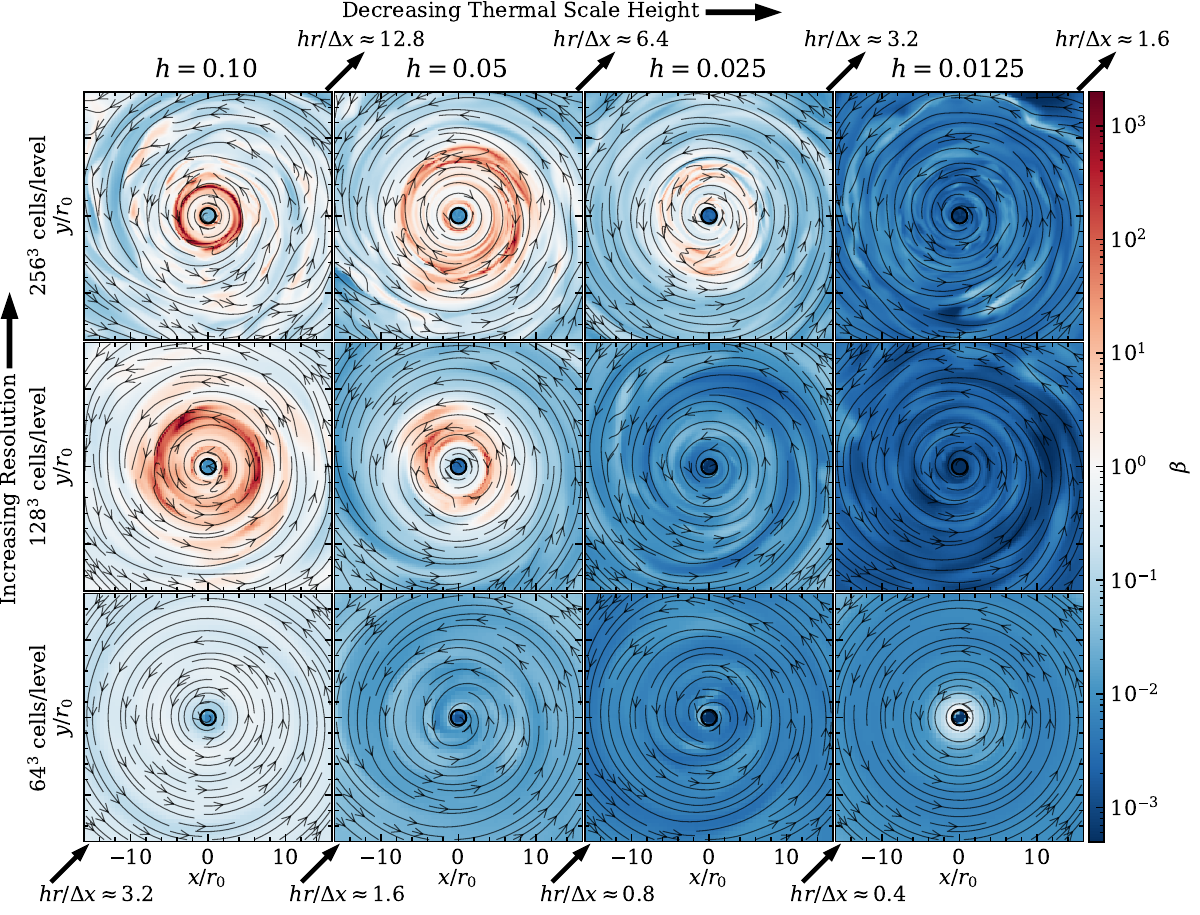}
\caption{Slices of the midplane plasma $\beta$ (i.e., $\beta$ evaluated in the $z=0$ plane) with magnetic field lines overlaid, shown for different thermal scale heights and numerical resolutions in the quasi-steady state. The snapshots are taken at $t = 15\,\mathrm{k}\,t_0$ for the run with $256^3$ cells per level and $t = 30\,\mathrm{k}\,t_0$ for the other runs. Panels are arranged such that, from left to right, the thermal scale height decreases, while from bottom to top the resolution increases (in terms of the number of cells per dimension at each refinement level). The diagonal panels from lower left to upper right correspond to constant relative resolution, measured in terms of the number of cells per thermal scale height. We find that the midplane $\beta \sim 1$ when the thermal scale height is sufficiently resolved, i.e., $hr/\Delta x \gtrsim 4$, whereas $\beta \ll 1$ when the scale height is under-resolved.
\label{fig:all_beta_slice}}
\end{figure*}

\begin{figure*}[ht]
\centering
\includegraphics[width=\linewidth]{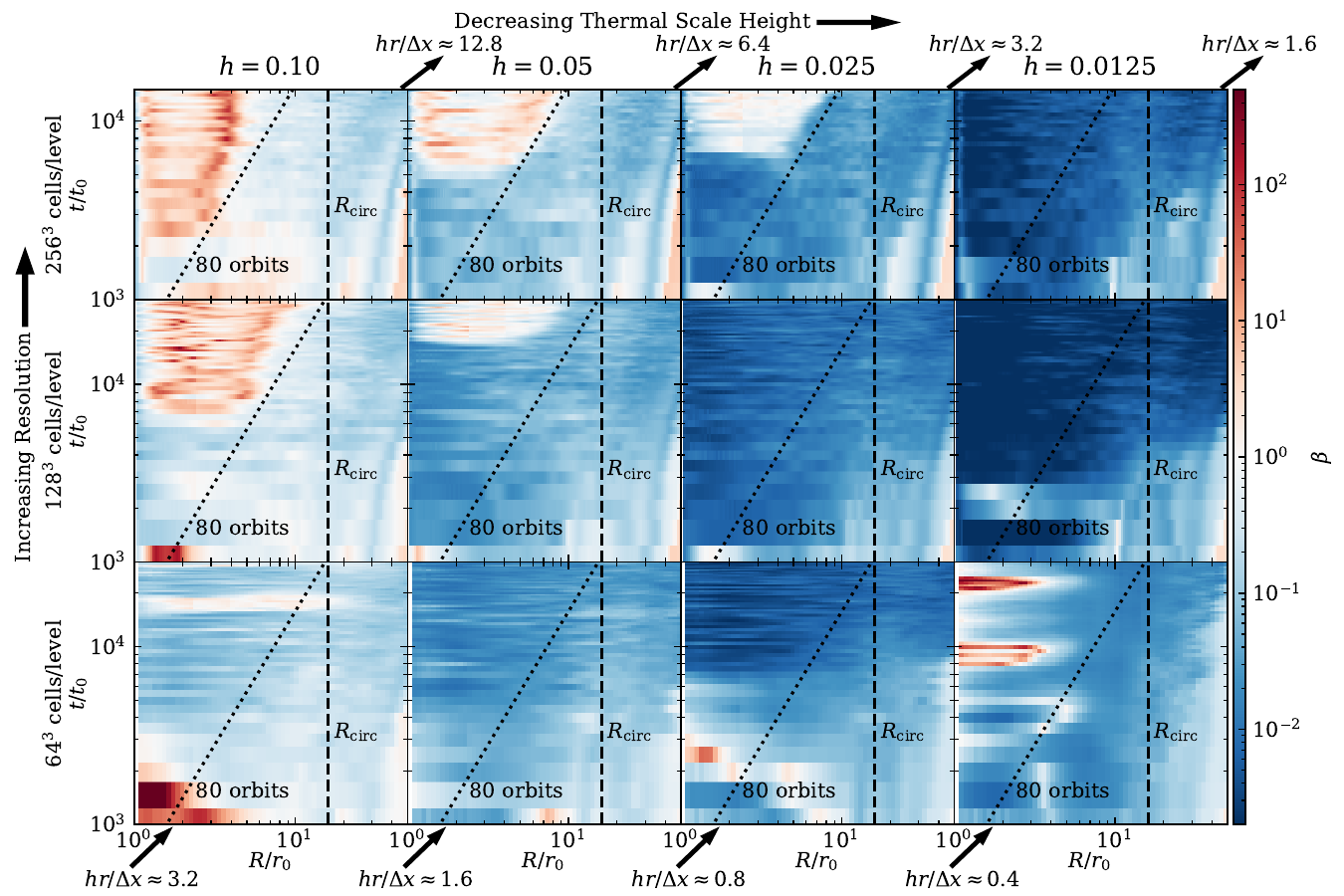}
\caption{Space-time diagrams of azimuthally averaged midplane $\beta$ for the same set of runs as in \figu\ref{fig:all_beta_slice}, with panels arranged in the same order. The black dashed lines mark the circularization radius $R_\mathrm{circ}=16\,r_0$. The black dotted lines mark the time when the evolution time is $80$ orbital periods at the corresponding radius. The diagonal panels from lower left to upper right show the same relative resolution in terms of cells per thermal scale height. The $H_\mathrm{th}$-unresolved cases maintain $\beta\ll 1$ while the $H_\mathrm{th}$-resolved cases gradually transition to a disk with midplane $\beta\sim1$ after $\sim 50-80$ orbits at the corresponding radius.
\label{fig:all_beta}}
\end{figure*}

\begin{figure*}[ht]
\centering
\includegraphics[width=\linewidth]{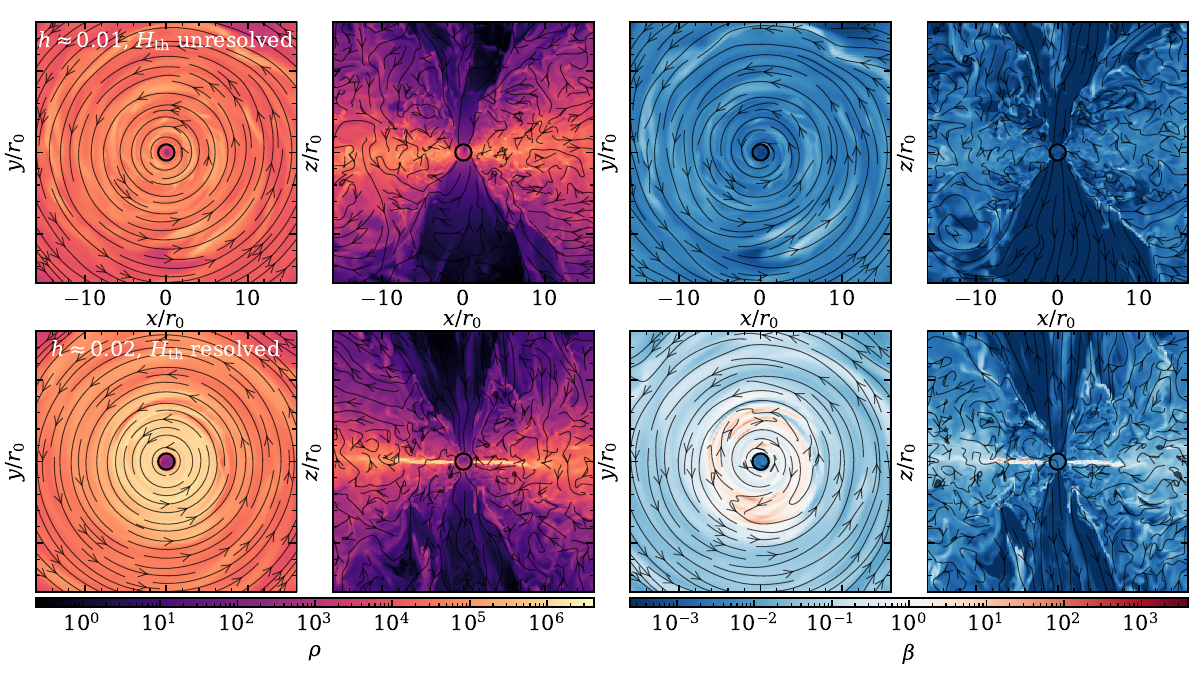}
\caption{Face-on and edge-on slices of (left) density with velocity streamlines and (right) plasma-$\beta$ with magnetic field lines at the end of the simulation. Top: a case with the thermal scale height ($h=0.0125$, model h01n256 at $t/t_0=1.5\times10^4$) unresolved. Bottom: a case with the thermal scale height marginally resolved ($h=0.025$, model h02n256 at $t/t_0=1.5\times10^4$). When $H_\mathrm{th}$ is unresolved, the disk is highly magnetized with $\beta\ll1$. When $H_\mathrm{th}$ is resolved, the disk is still $\beta\ll1$ outside the midplane, which is insensitive to resolution, but moderately magnetized close to the midplane with $\beta\sim1$ inside $R_\mathrm{circ}$.
\label{fig:beta_slice}}
\end{figure*}

\begin{figure*}[ht]
\centering
\includegraphics[width=0.49\linewidth]{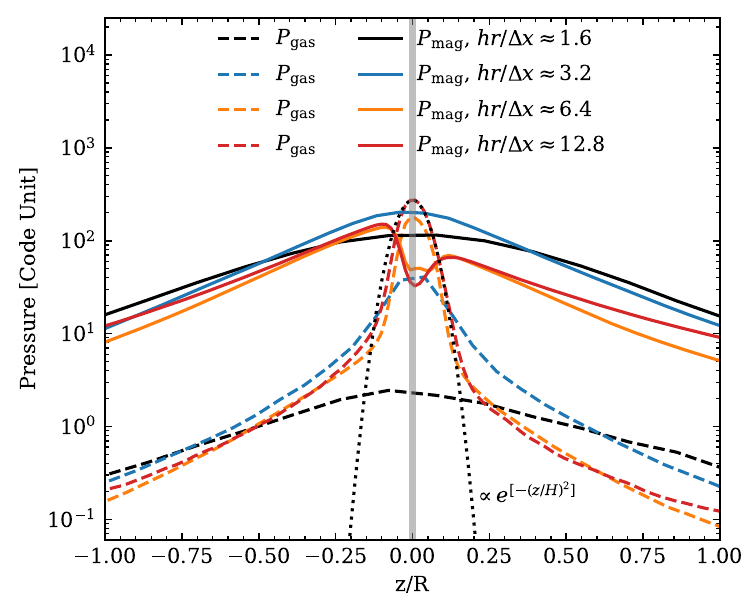}
\includegraphics[width=0.49\linewidth]{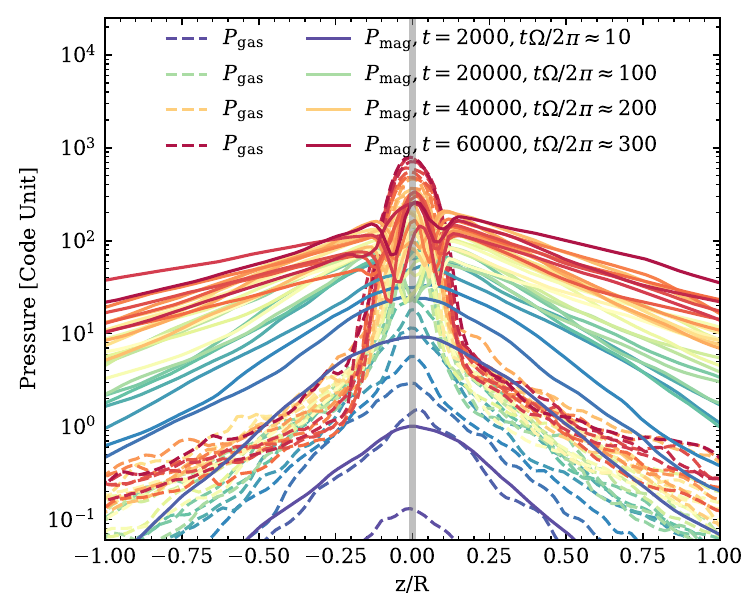}
\caption{Left: vertical profile of time and azimuthally averaged magnetic and thermal pressures at $R/r_0=10$ for different resolutions, i.e., models h05n64, h05n128, h05n256, and h05n512t3. The profile is averaged over $3\times10^4<t/t_0<3.3\times10^4$ for model h05n512t3 and $2.5\times10^4<t/t_0<3.0\times10^4$ for others. The black dotted line shows a Gaussian profile with $H=H_\mathrm{th}=\sqrt{2}hR$, and the vertical line marks the midplane, $z=0$. The thermal pressure in the midplane $P_\mathrm{gas}\ll P_\mathrm{mag}$ for the case with lowest resolution of $hr/\Delta x\approx 1.6$ but $P_\mathrm{gas}\sim P_\mathrm{mag}$ otherwise. Right: similar to left but for azimuthally averaged magnetic and thermal pressures as a function of time, from blue to red in equal time increments of $2000\,t_0\approx10$ orbits at $R/r_0=10$ for the fiducial simulation (model h05n256). For clarity, only four of these times are labeled. It takes several tens of orbits for the disk to collapse to the $\beta\sim1$ state.
\label{fig:pressure}}
\end{figure*}

\begin{figure*}[ht]
\centering
\includegraphics[width=\linewidth]{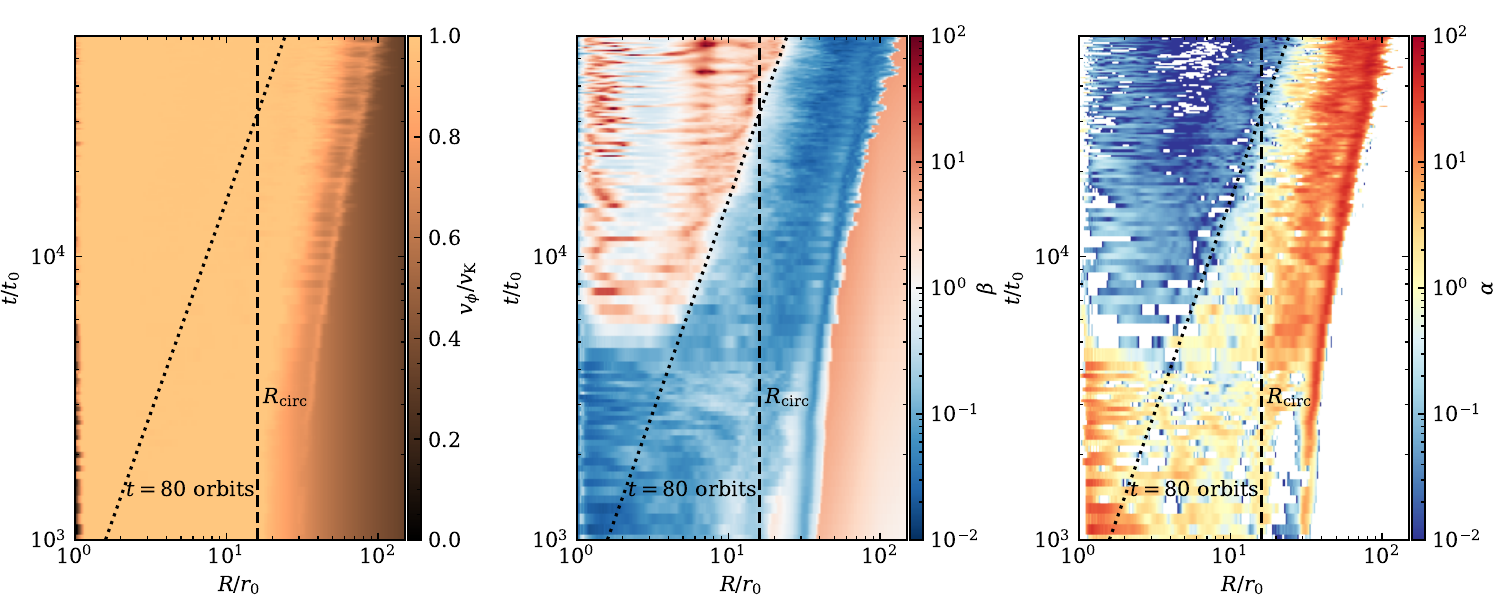}
\caption{Space-time diagram of azimuthally averaged midplane $v_\phi/v_\mathrm{K}$ (left), $\beta$ (middle), and $\alpha$ (right) for the fiducial run (model h05n256). The black dashed lines mark the circularization radius $R_\mathrm{circ}=16\,r_0$. The black dotted lines mark the time when the evolution time is $80$ orbital periods at the corresponding radius. The gas first forms a strongly magnetized ($\beta\ll1$), rapidly accreting ($\alpha\gg1$), Keplerian ($v_\phi\approx v_\mathrm{K}$) disk, then gradually collapses to a less magnetized ($\beta\sim1$ and $\alpha\gtrsim0.1$) state outwards over $\sim 50-80$ orbits. Outside $\sim R_\mathrm{circ}$, the disk is still in the $\beta\ll1$ and $\alpha\gg1$ state.
\label{fig:hist}}
\end{figure*}

\begin{figure*}[ht]
\centering
\includegraphics[width=0.49\linewidth]{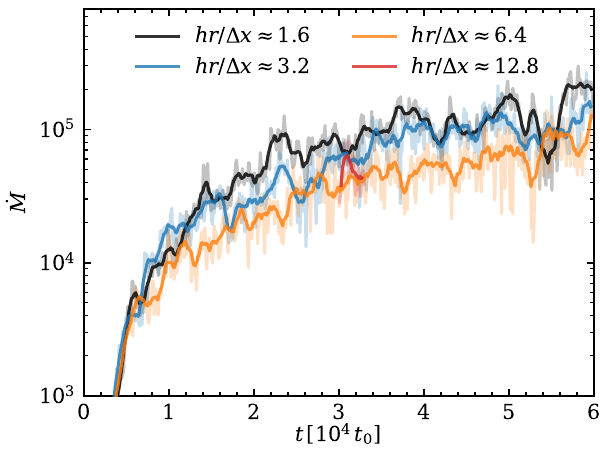}
\includegraphics[width=0.49\linewidth]{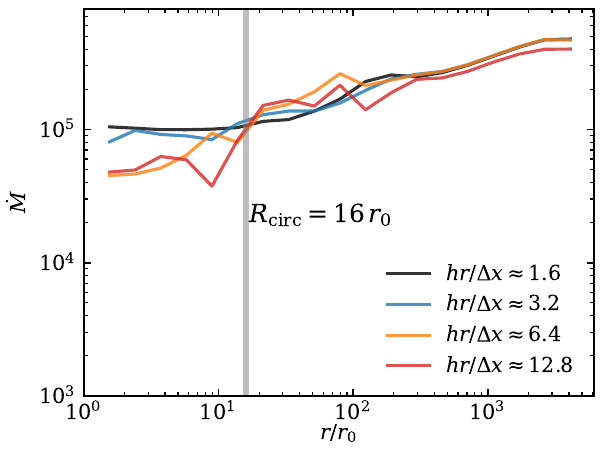}
\caption{Left: accretion rate as a function of time for the run with the same thermal scale height ($h=0.05$) but different resolutions (models h05n64, h05n128, h05n256, and h05n512t3). The curves are smoothed for visualization, with the original unsmoothed data shown as semi-transparent lines. The dependence of the accretion rate on resolution is relatively weak, while the dependence of midplane gas pressure on resolution is strong (see \figu\ref{fig:pressure}). Right: accretion rate as a function of radius averaged over the end of the simulation ($3.1\times10^4<t/t_0<3.3\times10^4$ for model h05n512t3 and $3\times10^4<t/t_0<4\times10^4$ for others). The vertical line marks the characteristic circularization scale $R_\mathrm{circ}=16\,r_0$. The profile of mass accretion rate is overall flat, though it slightly increases with increasing radius since the simulation time is not much longer than the free-fall time at large radii ($\sim 10^3\,r_0$).
\label{fig:mdot}}
\end{figure*}

\begin{figure*}[ht]
\centering
\includegraphics[width=0.98\linewidth]{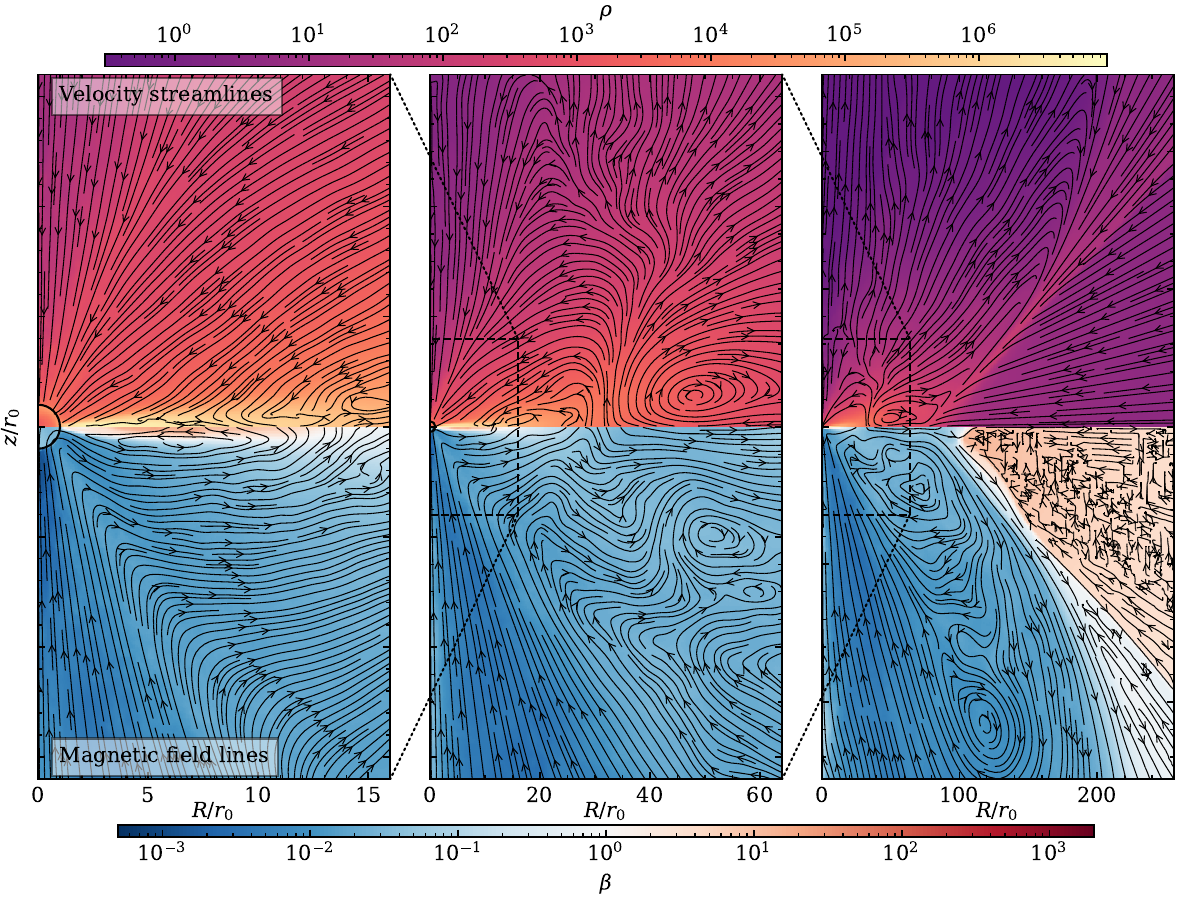}
\caption{Time and azimuthally averaged images of density (upper) and plasma-$\beta$ (lower) of the accretion disk from small (left, $R/r_0<16$) to large (right, $R/r_0<256$) scales for a model with the thermal scale height resolved (model h05n512t3 averaged over $3\times10^4<t/t_0<3.3\times10^4$). The gas inflows from large scales to the disk on small scales with outflows in the polar region. A mean radial field was generated due to the inflow. The radial field finally transitions to a vertical field and extends outward from two sides of the disk, forming large-scale loops with zero net vertical flux. The disk is magnetized with midplane $\beta\sim1$ and $\beta\ll1$ everywhere outside the midplane.
\label{fig:Rz}}
\end{figure*}

\begin{figure*}[ht]
\centering
\includegraphics[width=0.49\linewidth]{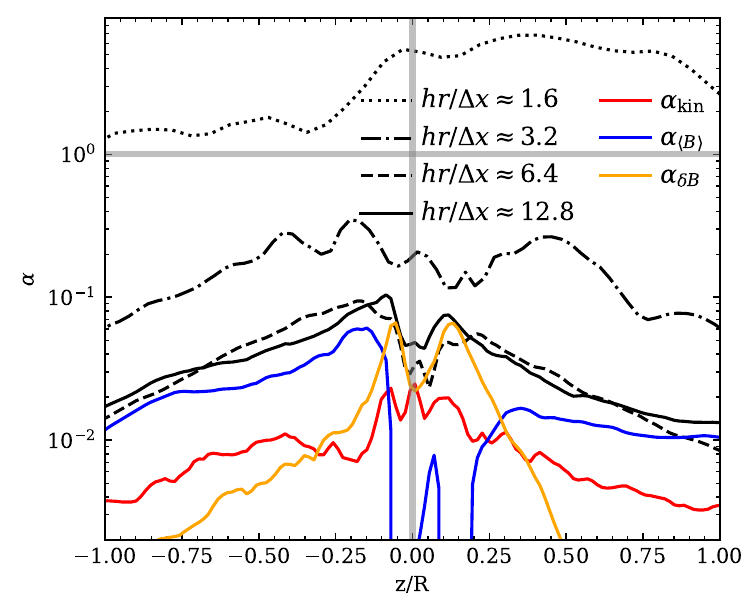}
\includegraphics[width=0.49\linewidth]{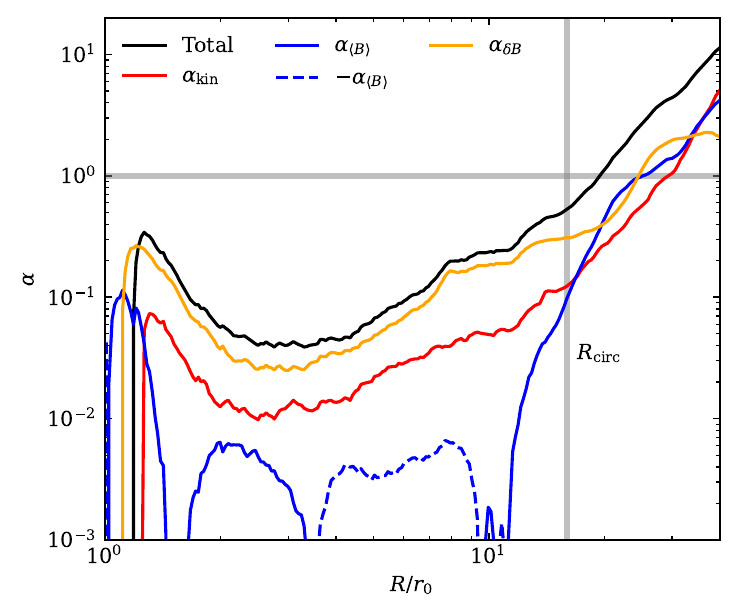}
\caption{Left: vertical profile of time and azimuthally averaged $\alpha$ (defined in \eq\ref{eq:alpha}) measured at $R/r_0=10$ for different resolutions (the same models and time range as in \figu\ref{fig:pressure}) and its components for the highest resolution run. The disk is accreting efficiently with $\alpha\sim 0.1$. The accretion is dominated by the fluctuating Maxwell stresses in the midplane and by the mean stresses outside $|z/H_\mathrm{th}|\gtrsim 2$. The parameter $\alpha\gtrsim 1$ if the thermal scale height is not resolved. Right: radial profiles of time and azimuthally averaged $\alpha$ ($|z/R|<0.2$) and its three components for model h05n512t3. The disk shows $\alpha\sim0.1$ with a weak dependence on radius inside $R_\mathrm{circ}$. The turbulent magnetic part $\alpha_{\delta B}$ dominates the angular momentum transfer. Between $R_\mathrm{circ}$ and the approximate outer edge of the rotationally supported disk, $R_\mathrm{edge} \simeq 40\,r_0$, there is stronger angular momentum transfer with $\alpha\gtrsim1$ dominating by both the mean stress $\alpha_\mathrm{\langle B\rangle}$ and the fluctuating stress $\alpha_\mathrm{\delta B}$.
\label{fig:alpha}}
\end{figure*}

\begin{figure}[ht]
\centering
\includegraphics[width=\linewidth]{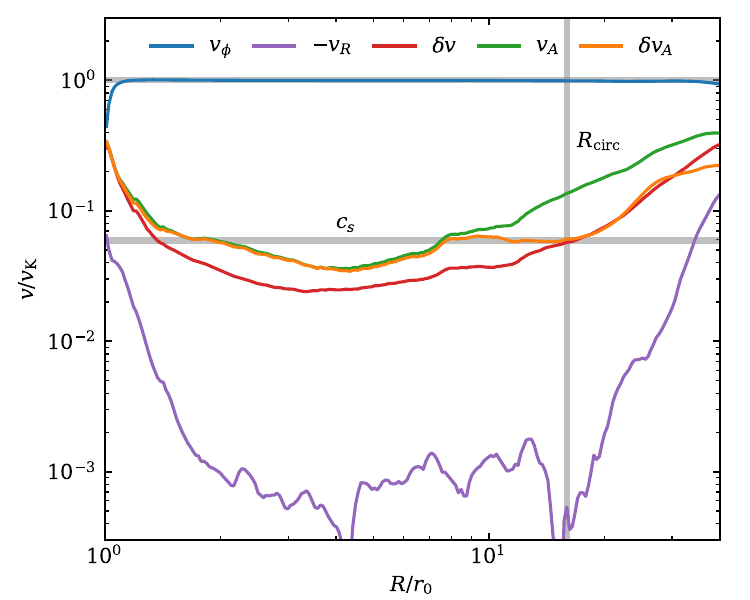}
\caption{Radial profiles of mass-weighted, time and azimuthally averaged rotational velocity, radial velocity, turbulent velocity, mean Alfv\'en velocity, and turbulent Alfv\'en velocity of the disk for $|z/R|<0.2$ as a function of cylindrical radius for the model h05n512t3. The disk is rotationally supported within $\sim 40\,r_0$. Radial advection is important outside $R_\mathrm{circ}$ with mean radial velocity $-v_\mathrm{R}/v_\mathrm{K}\sim 10^{-1}$. Within $R_\mathrm{circ}$, the turbulence is transonic and trans-Alfv\'enic with $\delta v\sim \delta v_A\sim v_A\sim c_s$.
\label{fig:vel_radial}}
\end{figure}

\begin{figure*}[ht]
\centering
\includegraphics[width=0.49\linewidth]{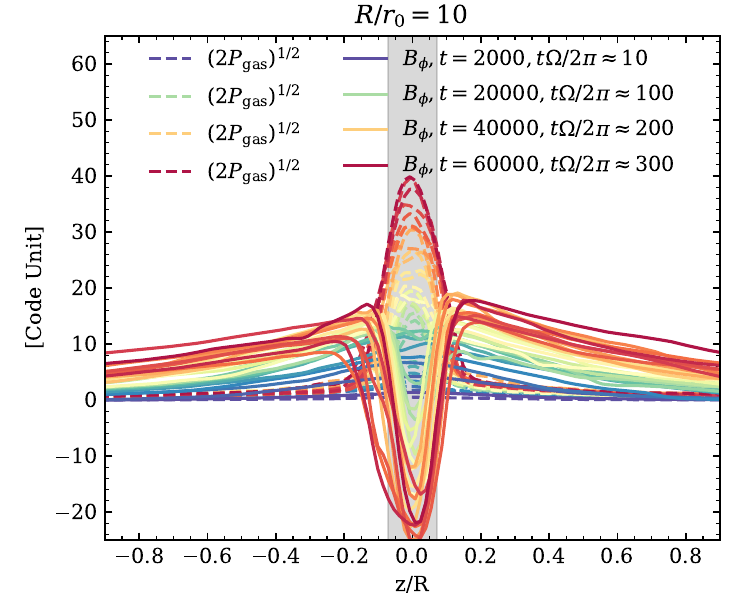}
\includegraphics[width=0.49\linewidth]{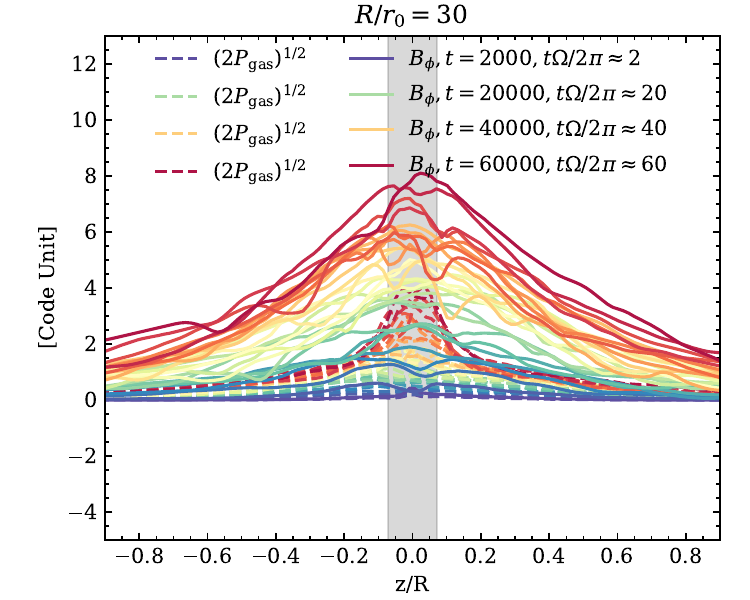}\\
\includegraphics[width=0.49\linewidth]{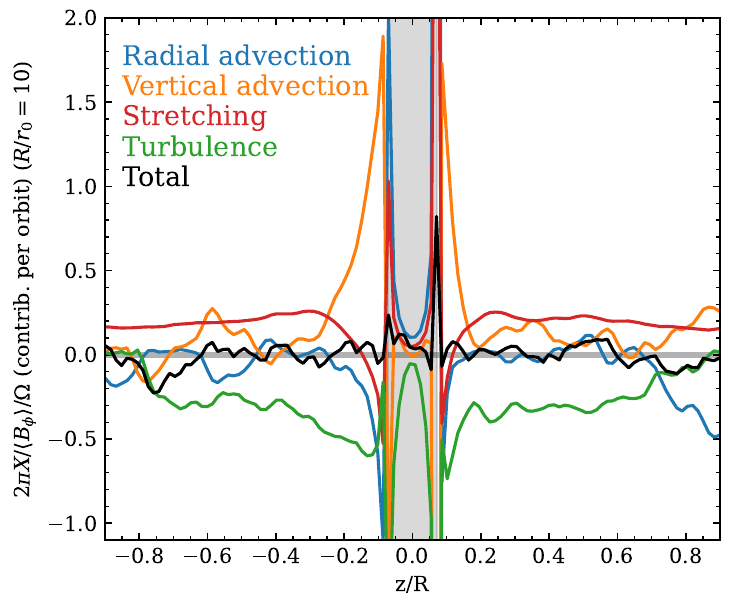}
\includegraphics[width=0.49\linewidth]{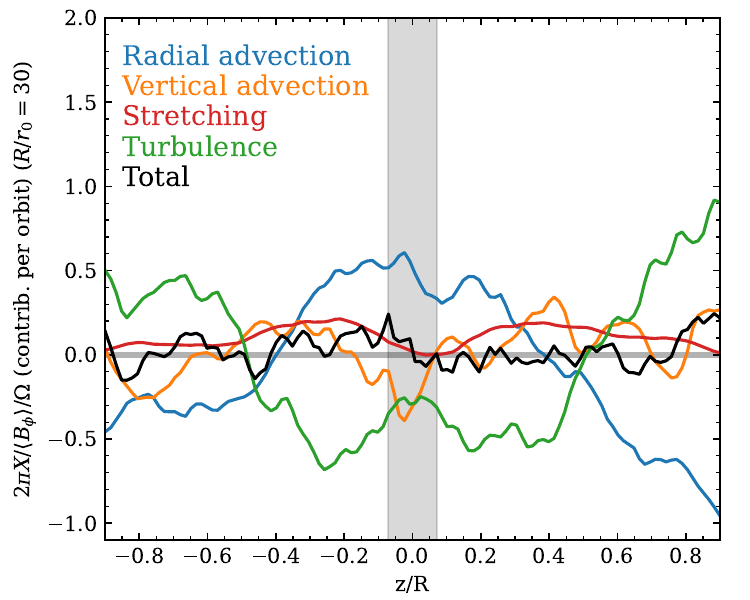}
\caption{Top: vertical profile of azimuthal magnetic field $B_\phi$ and $(2P_\mathrm{gas})^{1/2}$ for comparison as a function of time, from blue to red in equal time increments of $2000\,t_0$ at $R/r_0=10$ (left) and $30$ (right). For clarity, only four of these times are labeled. The gray strips mark the thermal scale height $H_\mathrm{th}/R=\sqrt{2}h$. 
Bottom: contribution of each term, denoted by $X$, in the induction equations~\eq\ref{eq:dynamo_phi} averaged over $3\times 10^4 < t/t_0 < 6\times10^4$. Total includes all the terms on the right-hand side of the averaged induction equation. Each term is normalized by $\langle B_\phi \rangle\Omega(R)/(2\pi)$ such that a value of $1$ implies that the particular term would grow or decay $\langle B_\phi\rangle$ at a given $z$ by 100\% over one orbit. The gray strips mark the thermal scale height $H_\mathrm{th}/R=\sqrt{2}h$. At $R/r_0=10$ where midplane $\beta\sim1$ and the field flips, the fields are replenished and expelled over several orbit timescales, with strong flux creation by shearing of radial field balanced by its diffusion via turbulence. The radial advection and compression are less important.
At $R/r_0=30$ where $\beta\ll1$, the radial advection of the field $B_\phi$ replenishes the field in the midplane quickly.
\label{fig:dynamo}}
\end{figure*}

\begin{figure*}[ht]
\centering
\includegraphics[width=\linewidth]{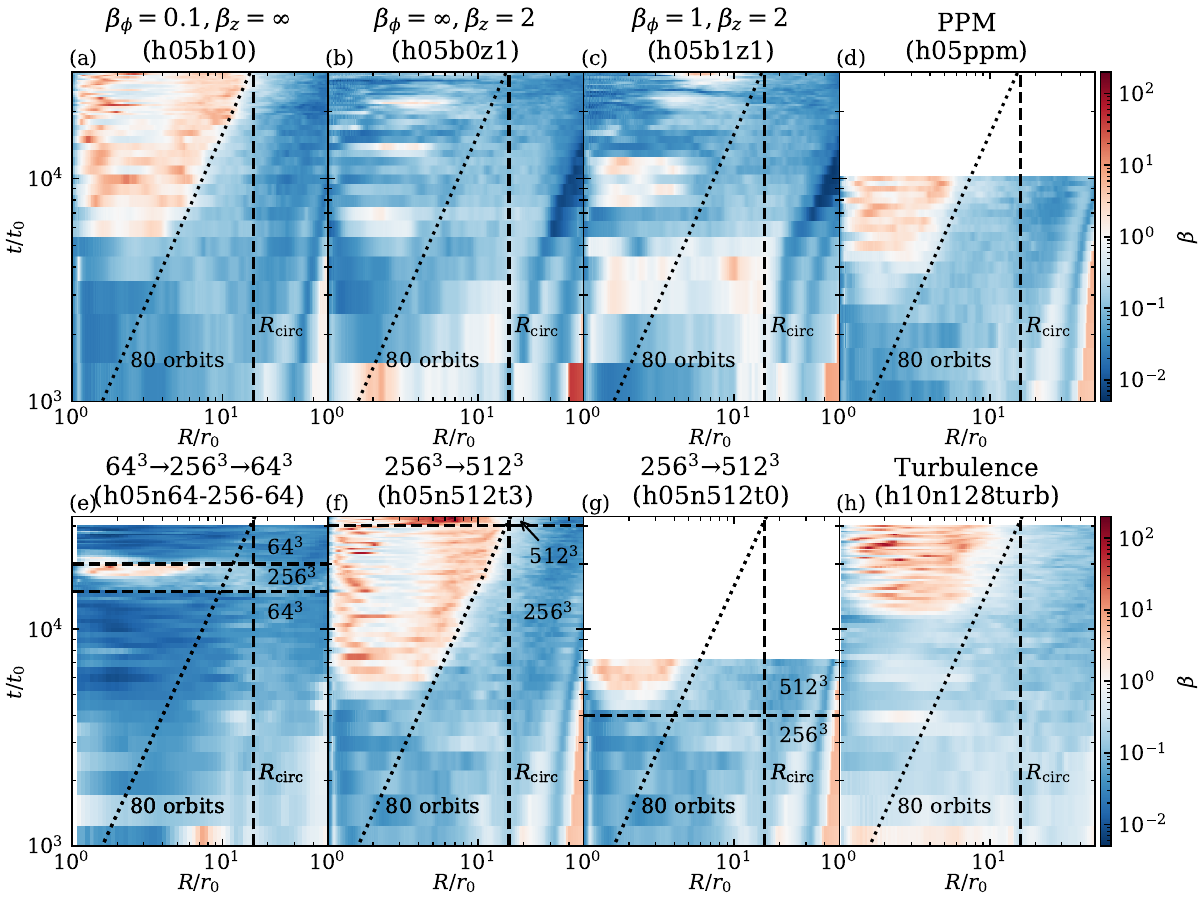}
\caption{Space-time diagram of azimuthally averaged midplane $\beta$, similar to \figu\ref{fig:all_beta}, but for miscellaneous tests. (a) Stronger initial magnetic field strength shows a similar transition to the $\beta\sim1$ disk within $\sim R_\mathrm{circ}$ over $\sim 80$ orbits. (b) Disk with initially purely vertical magnetic field sustains a strongly magnetized ($\beta<1$) state. (c) Disk with both toroidal and vertical field shows a similar strongly magnetized ($\beta<1$) state. (d) Higher-order PPM reconstruction method confirms the disk state transition similar to the model using PLM. (e) An initially $H_\mathrm{th}$-unresolved disk gradually collapses to $\beta\sim1$ after two levels of mesh refinement and returns to $\beta\ll1$ after derefinement. (f) The $\beta\sim1$ disk is maintained after doubling the resolution from $256^3$ to $512^3$ cells per level. (g) The initial collapse around $\sim 5\times10^3\,t_0$ in the fiducial run is reproduced in a higher-resolution restart. (h) A run with initial turbulence still undergoes collapse, albeit on a longer timescale (after $\sim 10^4\,t_0$).
\label{fig:beta_misc}}
\end{figure*}

\begin{figure}[ht]
\centering
\includegraphics[width=\linewidth]{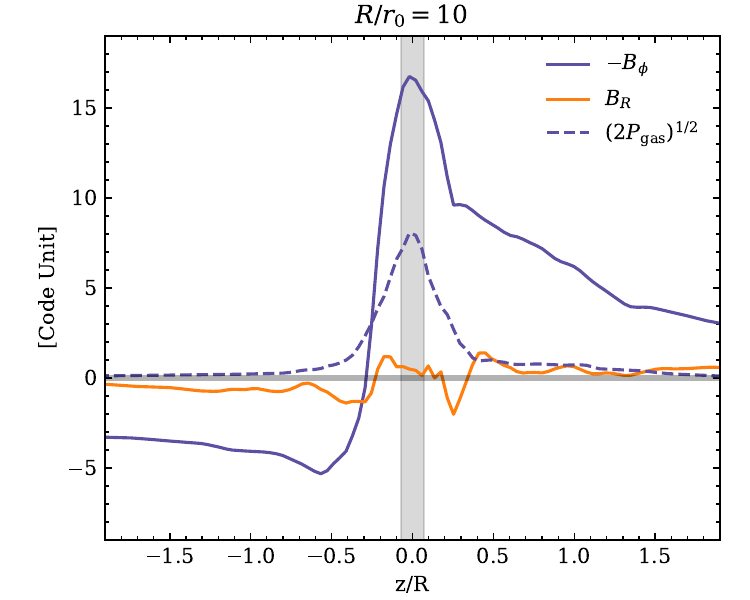}
\caption{Vertical profile of azimuthally averaged $B_\phi$, $B_R$, and $(2P_\mathrm{gas})^{1/2}$ at $R/r_0=10$ for the model initialized with vertical magnetic field (model h05b0z1) at the end of the simulation ($t/t_0=3\times10^4$). The gray strips mark the thermal scale height $H_\mathrm{th}/R=\sqrt{2}h$. The magnetic field dominates the midplane with a polarity flip near the midplane.
\label{fig:b0z1_vertical}}
\end{figure}

\section{Method} \label{sec:method}

\subsection{Physical Setup} \label{subsec:setup}
We solve the ideal MHD equations,
\begin{align}
    \frac{\partial \rho}{\partial t}+\nabla \cdot(\rho \boldsymbol{v}) &=0, \label{eq:mhd_mass_eq}\\
    \frac{\partial \rho \boldsymbol{v}}{\partial t}+\nabla \cdot\left(P_\mathrm{tot}\mathbf{I}+\rho \boldsymbol{v} \boldsymbol{v}-\boldsymbol{B} \boldsymbol{B}\right) &=\rho\boldsymbol{g}, \label{eq:mhd_momentum_eq}\\
    \frac{\partial E}{\partial t}+\nabla \cdot\left[\left(E+P_\mathrm{tot}\right) \boldsymbol{v}-\boldsymbol{B}\left(\boldsymbol{B} \cdot \boldsymbol{v}\right)\right] &=\rho\boldsymbol{g}\cdot\boldsymbol{v}, \label{eq:mhd_energy_eq}\\
    \frac{\partial \boldsymbol{B}}{\partial t}-\nabla \times(\boldsymbol{v} \times \boldsymbol{B}) &=0, \label{eq:mhd_induction_eq}
\end{align}
where $\rho$ is the gas density, $\boldsymbol{v}$ is the velocity, $P_\mathrm{tot}=P_\mathrm{gas}+P_\mathrm{mag}$ with $P_\mathrm{gas}=\rho T$ the gas pressure and $P_\mathrm{mag}=B^2/2=\boldsymbol{B}\cdot\boldsymbol{B}/2$ the magnetic pressure, $\boldsymbol{B}$ is the magnetic field, $\mathbf{I}$ is the identity matrix, $\boldsymbol{g}=-GM\boldsymbol{r}/r^3$ is the gravitational acceleration, with $M$ being the central mass, $E=E_\text{int}+\rho v^2/2+B^2/2$ is the total energy density with $E_\text{int}=P_\mathrm{gas}/(\gamma-1)$ the internal energy density where $\gamma=1.4$. The adiabatic index $\gamma$ is only used in the Riemann solver with sound speed $c_s=\sqrt{\gamma T}$. The equation of state we use is locally isothermal by fixing the temperature at all times to be the initial value $T=T_\mathrm{init}$ (described below). These equations are written in units such that the magnetic permeability $\mu_{m}=1$. The Alfv\'en speed is $v_A\equiv B/\sqrt{\rho}$ in such units.

The initial conditions for our simulations are a cloud of gas with constant density and angular momentum and a locally isothermal equation of state in which the temperature depends on cylindrical radius. Over time this gas cloud flows inwards, circularizes, and forms a disk with an initially strong toroidal magnetic field. More specifically, the initial gas density, temperature, and velocity are set using cylindrical coordinates $(R,\phi,z)$ by
\begin{align}
    \rho_\text{init}(R,z)&=
    \begin{cases}
      \rho_0 & \text{if $R\geq R_\text{circ}$},\\
      \rho_\text{floor}^{1-f}\rho_0^{f}, & \text{if $R_\text{smooth}\leq R < R_\text{circ}$},\\
      \rho_\text{floor} & \text{otherwise},
    \end{cases} \\
    T_\text{init}(R,z)&=\max\left(h^2\frac{GM}{R},T_\text{inf}\right),\label{eq:ic_temp}\\
    v_{\phi,\text{init}}(R,z)&=
     \begin{cases}
      \frac{j_\text{circ}}{R} & \text{if $R\geq R_\text{circ}$},\\
      \frac{j_\text{circ}}{R}\left(\frac{R}{R_\text{circ}}\right)^{1/2} & \text{otherwise},
    \end{cases}
\end{align}
where  
\begin{equation}
    f=\frac{\ln{(R/R_\text{smooth})}}{\ln{(R_\text{circ}/R_\text{smooth})}},
\end{equation}
is a smoothing factor so that the density smoothly changes from $\rho_\mathrm{floor}$ at $R_\mathrm{smooth}$ to $\rho_0$ at $R_\mathrm{circ}$.
Here $h$ is a free parameter controlling the thermal scale height $H_\mathrm{th}=\sqrt{2}hR$, $T_\text{inf}$ is the temperature at infinity, and $j_\mathrm{circ}=j_\mathrm{K}(R_\mathrm{circ})=\sqrt{GMR_\mathrm{circ}}$ is the Keplerian angular momentum at the specified circularization radius $R_\mathrm{circ}$. The corresponding Keplerian velocity is defined as $v_\mathrm{K}(R)\equiv\sqrt{GM/R}$ and the Keplerian angular velocity is $\Omega(R)\equiv\sqrt{GM/R^3}$. The temperature is set in order to enforce a thermal scale height in the roughly Keplerian portion of the flow that is proportional to the cylindrical radius.

The magnetic field is initialized by numerically calculating $\boldsymbol{B}=\nabla\times\boldsymbol{A}$, where the vector potential is $\boldsymbol{A}=A_{z,\text{init}}\hat{\boldsymbol{z}}$ with 
\begin{align}
    A_{z,\text{init}}&=
    \begin{cases}
      -A_\text{0}(R-R_\text{circ}) & \text{if $R\geq R_\text{circ}$},\\
      0 & \text{otherwise},
    \end{cases}
\end{align}
to construct a purely toroidal field outside $R_\mathrm{circ}$. The vector potential normalization $A_0$ is chosen to set a specified mean $\beta_\phi \equiv {\int P_\text{gas}dV}/{\int B_\phi^2/2dV}$ at large radii in the initial condition. We also run simulations with net vertical flux in addition to the toroidal field. In those runs, we add a uniform vertical magnetic field $B_z$ across the full domain with a specified mean $\beta_z\equiv \int P_\mathrm{gas}dV / \int B_z^2/2dV$. Note that both $\beta_\phi$ and $\beta_z$ are volume-integrated initial-condition parameters, to be distinguished from the local plasma $\beta\equiv P_\mathrm{gas}/P_\mathrm{mag}$ used throughout the analysis.

In all runs, we use $G=M=1$ and set the code length unit $r_0=1$. The corresponding time unit is $t_0 = (r_0^3/GM)^{1/2}=1$. We fix the outer boundary conditions to be the initial conditions. The fixed outer boundary conditions have negligible effects on the accretion flow at smaller radii because of the large separation of spatial and time scales. For the inner boundary ($r<r_\mathrm{in}\equiv r_0=1$), we adopt a vacuum sink by evacuating the gas within the region and resetting a fixed density $\rho_\mathrm{sink}=\rho_\mathrm{floor}=10^{-4}\ll \rho_0$, temperature $T_\mathrm{sink}=10^{-4}$, and zero velocity every time step. This acts as an absorbing inner boundary where gas entering the sink is removed from the computational domain. To avoid a large Alfv\'en speed within the sink, we apply a ceiling $v_{A,\mathrm{ceil}}=2$. We apply the ceiling by increasing the density after applying the density sink without changing the magnetic field.

\subsection{Numerical Method}\label{subsec:numerical_method}

We perform magnetohydrodynamic simulations using \athenak~\citep{Stone2026ApJS..283...27S}, a performance portable version of the \athenapp~\citep{Stone2020ApJS..249....4S} code implemented using the \kokkos\ library~\citep{Trott2021CSE....23e..10T}. \athenak\ provides a variety of reconstruction methods, Riemann solvers, and integrators for solving the MHD equations in a Cartesian grid. The adaptive mesh refinement (AMR) in \athenak\ allows us to flexibly achieve a high resolution and good performance over an extremely large dynamic range. In our simulations, we adopt the piecewise linear (PLM) reconstruction method, the Harten-Lax-van Leer discontinuities (HLLD) Riemann solver, and the 2nd order Runge-Kutta (RK2) time integrator to solve the MHD equations.

The simulations adopt a cubic domain spanning $[x_\mathrm{min},x_\mathrm{max}]\times[y_\mathrm{min},y_\mathrm{max}]\times[z_\mathrm{min},z_\mathrm{max}]=[-2^{12},2^{12}]^3\,r_0^3 = [-4096,4096]^3\,r_0^3$. In the fiducial run, the root grid is a cube of $N_x\times N_y\times N_z=256^3$ cells, which is divided into $4\times4\times4$ mesh blocks, with each mesh block being a cube of $64^3$ cells of size $\Delta x = 32\, r_0$. At each level of mesh refinement, we double the resolution by refining the central $2\times2\times2$ parent mesh blocks into $4\times4\times4$ child mesh blocks. We apply 9 levels of mesh refinement with the finest resolution of $\Delta x_\mathrm{min}=r_0/16$ covering a domain of $[-8,8]^3\,r_0^3$. Each level is resolved with $256^3$ cells. In this way, we achieve approximately uniform relative resolution across a wide range of spatial scales. For analysis, we adopt cylindrical coordinates $(R,\phi,z)$ and spherical coordinates $(r,\theta,\phi)$, sharing the same origin and $z$-axis (the rotation axis) as the Cartesian grid. The three systems are related by the standard transformations $R=\sqrt{x^2+y^2}$, $\phi=\mathrm{atan2}(y,x)$, $r=\sqrt{x^2+y^2+z^2}=\sqrt{R^2+z^2}$, and $\theta=\arccos(z/r)$, with the cylindrical and Cartesian $z$ coordinate shared and the midplane at $z=0$. Due to the nested mesh hierarchy, the effective resolution is inherently scale-dependent. We quantify this by measuring the resolution as a function of spherical radius (see Appendix~\ref{app:resolution}).

\subsection{Simulation Runs} \label{subsec:runs}

We fix the parameters of the initial conditions to be  $\rho_0=1,\,\rho_\text{floor}=10^{-4},\,R_\mathrm{circ}=16\,r_0,\,R_\mathrm{smooth}=8\,r_0,\,T_\text{inf}=10^{-5}$. We set $R_{\mathrm{circ}} = 16\,r_0$ such that the flow at the circularization radius can be followed for a sufficient number of orbital timescales within the computational domain. The parameter $R_{\mathrm{smooth}}= 8\,r_0$ ensures a smooth initial profile and avoids introducing strong gradients or numerical artifacts near the inner boundary. The parameter $T_{\mathrm{inf}}$ is chosen such that the gas remains sufficiently cold compared to the gravitational potential. For the fiducial run we set $h=0.05$ and the vector potential normalization $A_0$ such that the plasma is initially magnetized with $\beta_\phi = 1$ at large radii (note that this corresponds to an initial magnetic energy that is small compared to the kinetic energy of the free-falling gas that develops as the simulation progresses, because $T_{\rm init}/(GM/R) \propto h^2$). For the cases with vertical flux, we set $\beta_z=2$ such that the initial average magnetic energy along all three directions $x$, $y$, and $z$ is the same. This means that the initial toroidal field is $B_\phi=\sqrt{2}B_z$ for $R>R_\mathrm{circ}$. For $R<R_\mathrm{circ}$, the vertical $\beta$ is initially very small because the gas density is very low, but can increase significantly later due to the inflow from large radii. These initial conditions are asymptotically a uniform medium on large scales. For the fiducial runs, we do not add any perturbations to the initial conditions. The turbulence that emerges in the simulations is generated from machine-level noise.

After evolving the fiducial case for $3\times 10^4\,t_0$, we restart the simulations and double the resolution everywhere (model h05n512t3). This run thus has a resolution of $512^3$ cells per level with $\Delta x_\mathrm{min}=r_0/32$, and we run it for an extra $3\times10^3\,t_0$ for the convergence study. In addition, we explore a parameter space of $h\in[0.1,0.05,0.025,0.0125]$ and resolution of $[256^3, 128^3, 64^3]$ cells per mesh refinement level to understand how the results change with the resolution relative to the thermal scale height. The finest resolution in each case is $\Delta x_\mathrm{min}=[1/16,1/8,1/4]\,r_0$, respectively. Furthermore, we run a series of tests varying the initial magnetic field strength, structure, numerical method, and mesh refinement and derefinement, which are presented in \sect\ref{subsec:miscellaneous}. The main parameters of all the runs are summarized in \tab\ref{tab:runs}. The set of runs costs about 200,000 GPU hours in total.

\section{Results} \label{sec:results}

\figu\ref{fig:proj} illustrates the gas morphology of the highest resolution model h05n512t3 from large to small scales at the end of the simulation ($t/t_0=3.3\times10^4$). The material from large scales ($\gtrsim 100\,r_0$) inflows radially and circularizes around $R_\mathrm{circ}$ due to the initial angular momentum, forming a rotationally supported turbulent accretion disk. Below we present the results of the simulations with a focus on the disk within $\sim R_\mathrm{circ}$. We first describe the resolution-dependent magnetic transition of the disk, then analyze the structure, vertical collapse, angular momentum transport, and magnetic-field evolution of the fiducial model, and finally present robustness tests varying the initial field strength and geometry, numerical method, and mesh refinement.

Before presenting the diagnostics, we define the averaging and fluctuation conventions used in the analysis below.
For a quantity $A$, the azimuthal average at fixed $(R,z)$ is
\begin{equation}
    \langle A \rangle_\phi \equiv \frac{1}{2\pi} \int_0^{2\pi} A d\phi.
\end{equation}
When time averaging is also applied over the interval $[t_i,t_f]$, we use
\begin{equation}
    \langle A \rangle_t \equiv \frac{1}{2\pi(t_f-t_i)} \int_{t_i}^{t_f}\int_0^{2\pi} A d\phi d t,
\end{equation}
where $t_i$ and $t_f$ are initial and final times. Mass-weighted averages are defined using the same averaging operator,
\begin{equation}
    \langle A \rangle_m \equiv \frac{\langle \rho A \rangle}{\langle \rho \rangle}.
\end{equation}
The fluctuating part is defined relative to the relevant mean, $\delta A = A - \langle A\rangle$.
The mass accretion rate through a sphere of radius $r$ is calculated in spherical coordinates by 
\begin{equation}
    \dot{M}=-\int_0^\pi \int_0^{2\pi} \rho v_r r^2 \sin\theta d\theta d\phi.
\end{equation}

\subsection{Magnetic transition of the accretion disk}

\figu\ref{fig:all_beta_slice} summarizes the magnetic state of the disks with $z=0$ slices of plasma-$\beta$ (i.e., midplane $\beta$) when the system is in a quasi-steady state for a set of runs varying the thermal scale height and resolution. The diagonal panels from lower left to upper right show the same relative resolution in terms of number of cells per thermal scale height. There are two classes of solutions for midplane $\beta$ depending on the resolution relative to the thermal scale height. If the thermal scale height is not resolved, the disk is strongly magnetized with $\beta\ll1$. Instead, if the disk is resolved with $hr/\Delta x\gtrsim 3$ on average (i.e., $\gtrsim 4$ cells per $H_\mathrm{th}$), the disk begins to be moderately magnetized with $\beta\sim1$. The disks with the same relative resolution show similar behaviors. We note that the $\beta\sim1$ midplane is also very different from the traditional disk with ZNVF. Instead, it is similar to the low $\beta$ state in \citet{Squire2024rapidstronglymagnetizedaccretion}.

Space-time diagrams of the azimuthally averaged midplane plasma-$\beta$ for the same runs are shown in~\figu\ref{fig:all_beta}. The $H_\mathrm{th}$-unresolved cases show a persistent $\beta\ll 1$. The only exception is the case h01n64, where the disk transitions to a $\beta\sim1$ occasionally, likely because the resolution is too low ($hr/\Delta x \approx 0.4$) to resolve the accretion disk. If the thermal scale height is resolved, during the early stage of the evolution, the disk still shows a midplane $\beta\ll1$ which is similar to the unresolved case. However, it gradually transitions to a disk with midplane $\beta\sim1$ after several tens of orbits. The transition radius moves out in time over $\sim 50-80$ orbits at the corresponding radius. \figu\ref{fig:beta_slice} more clearly shows the difference between the resolved and unresolved cases (h01n256 and h02n256) by face-on and edge-on slices of density and plasma-$\beta$ at a later stage of the evolution. When the thermal scale height is resolved, the disk is much thinner with scale height $H\sim H_\mathrm{th}$, more dense, moderately magnetized with $\beta\sim1$, and less turbulent in the midplane inside $R_\mathrm{circ}$. The disk still maintains the diffuse, strongly magnetized ($\beta\ll1$) state outside the midplane, which is insensitive to resolution.

\figu\ref{fig:pressure} more quantitatively shows the time and azimuthally averaged vertical profiles of magnetic pressure compared with thermal pressure for models with the same thermal scale height ($h=0.05$) but different resolutions (models h05n64, h05n128, h05n256, and h05n512t3). When the thermal scale height is unresolved (model h05n64 with $hr/\Delta x\sim1.6$), the disk is highly magnetized everywhere with $P_\mathrm{gas}\ll P_\mathrm{mag}$ in the midplane. However, when the thermal scale height is resolved with $hr/\Delta x\gtrsim 3.2$, the disk is much less magnetized with $P_\mathrm{gas}\sim P_\mathrm{mag}$ ($\beta\sim1$) in the midplane. When we further increase the resolution to $hr/\Delta x\sim 6.4$ or $12.8$, the profiles stay qualitatively similar with a slightly higher $P_\mathrm{gas}$ in the midplane. 
With increasing resolution, the magnetic pressure changes moderately, but the thermal pressure (and thus density since the disk is locally isothermal) in the midplane grows significantly as the thermal scale height transitions from unresolved to resolved; at high resolution, the gas pressure approaches a Gaussian profile near the midplane, indicating a vertical equilibrium in which gravity is balanced by thermal pressure. Once the gas pressure is important near the midplane, the magnetic pressure has a small ``dip'' near the midplane. We note that, when the thermal scale height is not resolved, the gas pressure profile near the midplane (and hence density since the gas is locally isothermal) is closer to a power-law than Gaussian.
These results are similar to the local shearing box simulations of \citet{Squire2024rapidstronglymagnetizedaccretion}.

\subsection{The structure of the accretion disk}

Here we focus on the details of the structure and evolution of the accretion disk in the fiducial run (model h05n256, which has a thermal scale height that is small but resolved). Space-time diagrams of the azimuthally averaged midplane $v_\phi/v_\mathrm{K}$, $\beta$, and $\alpha$ for the fiducial run in \figu\ref{fig:hist} illustrate the basic evolution and properties of the system. The gas first inflows radially due to the lack of support from rotation and pressure. Soon the gas inflow forms a rotationally supported ($v_\phi\approx v_\mathrm{K}$) disk for $R\lesssim 20\,r_0 \sim R_\mathrm{circ}$ due to the initial angular momentum; the outer radius with significant angular momentum slowly moves to larger radii over time as gas is accreted to small radii in the disk, transferring its angular momentum to larger radii. The disk is initially rapidly accreting with $\alpha\gg1$, strongly magnetized with $\beta\ll1$, and vertically supported by the magnetic pressure, similar to the disks found in \citet{Gaburov2012ApJ...758..103G}, \citet{Hopkins2024OJAp....7E..18H, Hopkins2024OJAp....7E..19H} and \citet{Guo2024arXiv240511711G}. After $\sim 5000\,t_0$, it gradually collapses vertically to a $\beta\sim1$ and $\alpha\sim0.1$ state; this vertical collapse is shown as a function of time also in \figu\ref{fig:pressure} above. This collapse proceeds inside out, occurring after a time of $\sim 50-80$ orbits at a given radius. Outside $\sim R_\mathrm{circ}$, the gas still has midplane $\beta\ll1$ at the end of the simulation. The magnetized disk at $\lesssim R_\mathrm{circ}$ exhibits a quasi-steady state after reaching the $\beta \sim 1$ state. The flow at larger radii is presumably affected by the circularization process, so we suspect that radii $\lesssim R_\mathrm{circ}$ are more representative of the ``generic'' toroidally dominated disk structure.

As shown in \figu~\ref{fig:mdot}, the accretion rate grows quickly in the early stage and reaches a quasi-steady state after~$2 \times 10^4\,t_0$. During the phase of rapidly increasing mass accretion rate the disk maintains $\alpha \gg 1$ and $\beta \ll 1$ while in the later stages when the mass accretion rate is only slowly increasing in time, the disk settles into the $\beta \sim 1$ state. Our qualitative interpretation of this result is that the magnetic flux is leaving the $\beta \ll 1$ disk due to buoyancy on a timescale of a few orbits (as shown in \sect~\ref{subsec:dynamo} below). When the mass accretion rate itself increases significantly over a few orbits, as it does at early times, the disk is able to maintain $\beta \ll 1$. But at later times when the accretion rate changes more slowly in time, the disk reaches its equilibrium quasi-steady structure with $\beta \sim 1$ and $\alpha \sim 0.1$. We also plot the accretion rate for other models with the same thermal scale height but different resolutions. Despite significant differences in disk geometry, the accretion rate as a function of both time and radius is very similar within a factor of $\sim 2$.

The time and azimuthally averaged density, plasma-$\beta$, velocity streamlines, and magnetic field lines on different scales are shown in \figu\ref{fig:Rz}. A magnetized disk forms within $R\lesssim 50\,r_0$. The disk is rotationally supported out to an approximate outer edge, which we denote by $R_\mathrm{edge}\approx 40\,r_0$. The outer part of the disk ($R\gtrsim R_\mathrm{circ}$) is strongly magnetized with $\beta\ll 1$. It transitions to a less magnetized disk with midplane $\beta\sim1$ for $R\lesssim R_\mathrm{circ}$, though keeps $\beta\ll1$ outside the midplane.
The gas inflows from large radii to the disk on small scales, with outflows in the polar region on large scales. The gas flows out in the $\beta\sim1$ midplane on small scales. A spatially coherent mean radial magnetic field is generated by flux freezing in the inflow. The radial field finally transitions to a vertical field at small radii near the sink, and extends outward from two sides of the disk, forming large-scale loops. There is still zero net vertical flux.

\subsection{Collapse towards the midplane}\label{subsec:collapse}

We find that resolving or not resolving the gas thermal scale height changes the accretion flow structure and dynamics significantly. When the gas thermal scale height is resolved, the midplane is a gas-pressure-dominated thin disk with density scale height $H_\rho\sim H_\mathrm{th}$, $\beta \sim 1$--$10$, and $\alpha \sim 0.1$. When the gas thermal scale height is not resolved, the midplane remains magnetically supported with density scale height $H_\rho > H_{\rm th}$, $\beta \ll 1$ and $\alpha \gg 1$. This strong dependence on the resolution relative to $H_{\rm th}$ is seen in \figus\ref{fig:all_beta_slice}, \ref{fig:all_beta}, \ref{fig:beta_slice}, and \ref{fig:pressure}.

Interestingly, however, the approach to a gas-pressure dominated midplane at high resolution occurs relatively slowly: the right panel of \figu\ref{fig:pressure} shows the vertical profile of gas pressure and magnetic pressure at $R/r_0=10$ as a function of time from 10 orbits to 150 orbits. When the disk first forms, the midplane is magnetically dominated (see also \figu\ref{fig:hist}) with $P_\mathrm{mag}\gg P_\mathrm{gas}$. It takes several tens of orbits for the gas to gradually collapse to the midplane and saturate to a state that is thermally dominated with $P_\mathrm{mag}\sim P_\mathrm{gas}$. This feature is also consistent with local shearing box simulations \citep{Squire2024rapidstronglymagnetizedaccretion}.

\subsection{Turbulence and angular momentum transfer}

To better diagnose the structure and angular momentum transfer of the accretion disk, we consider the stresses given by
\begin{equation}
    \boldsymbol{\Pi}_\mathrm{tot}=\underbrace{\rho\boldsymbol{v}\boldsymbol{v}}_{\boldsymbol{\Pi}_\mathrm{kin}} + \underbrace{P_\mathrm{gas}\boldsymbol{\mathrm{I}}}_{\boldsymbol{\Pi}_\mathrm{th}} + \underbrace{\bigg(\overbrace{\frac{\boldsymbol{B}\cdot \boldsymbol{B}}{2}\boldsymbol{\mathrm{I}}}^{\boldsymbol{\Pi}_\mathrm{pmag}}-\overbrace{\boldsymbol{B}\boldsymbol{B}}^{\boldsymbol{\Pi}_\mathrm{tension}}\bigg)}_{\boldsymbol{\Pi}_\mathrm{mag}}.\\
\end{equation}
The kinetic stresses can be further separated into $\boldsymbol{\Pi}_\mathrm{kin}=\bar{\boldsymbol{\Pi}}_\mathrm{kin}+\delta\boldsymbol{\Pi}_\mathrm{kin}$ where the mean part $\bar{\boldsymbol{\Pi}}_\mathrm{kin}=\rho\langle\boldsymbol{v}\rangle\langle\boldsymbol{v}\rangle$ with $\langle\cdot\rangle$ denoting time and angle average and the fluctuating part (the Reynolds stress) $\delta\boldsymbol{\Pi}_\mathrm{kin}=\rho\delta\boldsymbol{v}\delta\boldsymbol{v}$ with the fluctuating velocities $\delta \boldsymbol{v} = \boldsymbol{v} - \langle \boldsymbol{v}\rangle$. Similarly, the magnetic (Maxwell) stresses $\boldsymbol{\Pi}_\mathrm{mag}=\bar{\boldsymbol{\Pi}}_\mathrm{mag}+\delta\boldsymbol{\Pi}_\mathrm{mag}$ where $\bar{\boldsymbol{\Pi}}_\mathrm{mag}=|\langle\boldsymbol{B}\rangle|^2\boldsymbol{\mathrm{I}}/2-\langle\boldsymbol{B}\rangle\langle\boldsymbol{B}\rangle$ and $\delta\boldsymbol{\Pi}_\mathrm{mag}=|\delta\boldsymbol{B}|^2\boldsymbol{\mathrm{I}}/2-\delta\boldsymbol{B}\delta\boldsymbol{B}$ with fluctuating magnetic field $\delta \boldsymbol{B} = \boldsymbol{B} - \langle \boldsymbol{B}\rangle$. Then the $\alpha$ parameter is defined by
\begin{equation}
    \alpha = \frac{\delta \boldsymbol{\Pi}^{R\phi}_\mathrm{kin} + \bar{\boldsymbol{\Pi}}^{R\phi}_\mathrm{mag} + \delta\boldsymbol{\Pi}^{R\phi}_\mathrm{mag}}{\langle P_\mathrm{gas, midplane}\rangle}=\alpha_\mathrm{kin}+\alpha_{\langle B \rangle}+\alpha_{\delta B}\,
    \label{eq:alpha}
\end{equation}
where $P_\mathrm{gas, midplane}$ is the time and angle-averaged midplane gas pressure.
With this sign convention, positive $\alpha$ corresponds to outward angular momentum transport, while a negative component corresponds to inward transport by that stress component.
\figu\ref{fig:alpha} plots the vertical profiles of time and azimuthally averaged $\alpha$ and its components: $\alpha \sim 0.1$ with the accretion driven by both the mean and the fluctuating magnetic stresses. The fluctuating magnetic stresses dominate in the midplane ($|z/R|<0.1$), while the mean field dominates elsewhere.
This is similar to the local shearing box simulations \citep{Squire2024rapidstronglymagnetizedaccretion}.
The right panel of \figu\ref{fig:alpha} plots the radial profile of $\alpha$ and its components for the model h05n512t3. 
On all scales within $R_\mathrm{circ}$ we find $\alpha\sim0.1$ with a weak dependence on radius. Between $R_\mathrm{circ}$ and the outer edge of the disk $R_\mathrm{edge} \simeq 40\,r_0$, there is stronger angular momentum transfer with $\alpha\sim1$. It is dominated by both the mean field and the fluctuating field.

When the thermal scale height is not resolved, the turbulence is much larger with $\alpha\gg1$ (see \figus\ref{fig:alpha}); this is a consequence of similar magnetic field strengths but very different midplane gas densities and pressures in the two cases ($H_{\rm th}$ resolved and unresolved). In both the $H_{\rm th}$ resolved and unresolved regimes, however, the final mass accretion rate is quite similar within a factor of 2, as shown in \figu\ref{fig:mdot} -- this accretion rate is set by the free-fall rate of gas at large radii in our initial conditions. The resolved thermal scale height simulation has a smaller scale height and $\alpha$ at a given $\dot{M}$ relative to the unresolved thermal scale height simulation, leading to much higher midplane densities and pressures. We note that, the gas flows out for $|z|\lesssim H_\mathrm{th}$ and most of the accretion happens outside the midplane ($|z|\gtrsim H_\mathrm{th}$), as shown in \figu\ref{fig:Rz}.

\figu\ref{fig:vel_radial} compares the mass-weighted, time and azimuthally averaged mean and fluctuating velocities and Alfv\'en speeds in the midplane as a function of radius. These profiles quantitatively motivate our estimate $R_\mathrm{edge}\approx 40\,r_0$ for the outer edge of the rotationally supported disk. The turbulence within $R_\mathrm{circ}$ is essentially transonic and trans-Alfv\'enic with $\delta v^2\sim \delta v_A^2 \sim v_A^2 \sim c_s^2$. The turbulence is likely driven by a combination of the MRI and Parker instability~\citep[][]{Johansen2008AA...490..501J, Pessah2005ApJ...628..879P, Squire2024rapidstronglymagnetizedaccretion}. The radial velocity is $-v_\mathrm{R}/v_\mathrm{K}\sim 10^{-3}$ within $R_\mathrm{circ}$ but larger with $-v_\mathrm{R}/v_\mathrm{K}\sim 10^{-1}$ outside $R_\mathrm{circ}$, indicating the relative importance of radial advection on larger scales.

\subsection{Sustaining the magnetic field}\label{subsec:dynamo}

\citet{Squire2024rapidstronglymagnetizedaccretion} performed a detailed analysis of the dynamo in their local shearing box simulations of disks with strong toroidal fields. The results demonstrated that the net azimuthal field is sustained by a dynamo mechanism. In the shearing box, vertical turbulent diffusion and vertical outflows continuously remove azimuthal flux out of the top and bottom boundaries of the simulation. This loss is balanced by creation of azimuthal flux via shearing of the radial field. \citet{Squire2024rapidstronglymagnetizedaccretion}'s analysis of the origin of the radial field from this azimuthal flux was somewhat inconclusive and more work to understand this aspect of the dynamo is needed. Here we focus primarily on global effects not present in the shearing box, in particular the role of radial gas flow in advecting magnetic field in from larger radii.

A spatial and time average of the induction equation \ref{eq:mhd_induction_eq} yields 
\begin{equation}
    \frac{\partial \langle\boldsymbol{B}\rangle}{\partial t} = \nabla\times(\langle \boldsymbol{v}\rangle \times\langle\boldsymbol{B}\rangle)+\nabla\times\delta\boldsymbol{\mathcal{E}},
    \label{eq:dynamo}
\end{equation}
where $\langle\boldsymbol{v}\rangle$ is the time and azimuthally averaged mean flow, $\langle\boldsymbol{B}\rangle$ is the mean magnetic field, and the turbulent \textit{electromotive force} (EMF) $\delta\boldsymbol{\mathcal{E}} = \langle\delta\boldsymbol{v}\times\delta\boldsymbol{B}\rangle$, which is responsible for dynamo action, is nonzero due to the correlation between $\delta\boldsymbol{v}$ and $\delta\boldsymbol{B}$. We rewrite the induction equation as (note that $\nabla\cdot\boldsymbol{B}=0$)
\begin{equation}
    \frac{\partial \langle\boldsymbol{B}\rangle}{\partial t} = \overbrace{\underbrace{-\langle\boldsymbol{v}\rangle\cdot\nabla\langle\boldsymbol{B}\rangle}_{\text{advection}} \underbrace{-\langle\boldsymbol{B}\rangle\nabla\cdot\langle\boldsymbol{v}\rangle}_{\text{compression}}}^{\text{advection}}+\underbrace{\langle\boldsymbol{B}\rangle\cdot\nabla\langle\boldsymbol{v}\rangle}_{\text{stretching}}+\underbrace{\nabla\times\delta\boldsymbol{\mathcal{E}}}_{\text{turbulence}},
    \label{eq:dynamo_div_form}
\end{equation}
to show the advection (including compression as the flow is fully compressive), stretching, and turbulence terms. For the azimuthal field $B_\phi$, \eq\ref{eq:dynamo_div_form} becomes
\begin{equation}
\begin{aligned}
    \frac{\partial \langle B_\phi\rangle}{\partial t}=&\underbrace{-\left(\langle v_R\rangle \partial_R \langle B_\phi\rangle + \langle v_\phi\rangle \langle B_R\rangle/R + \langle B_\phi\rangle \partial_R \langle v_R\rangle\right)}_{\text{radial advection}} \\
    & \underbrace{- \left( \langle B_\phi\rangle \partial_z \langle v_z \rangle + \langle v_z\rangle \partial_z \langle B_\phi \rangle \right)}_{\text{vertical advection}} \\
    & + \underbrace{\left( \langle B_R\rangle \partial_R \langle v_\phi \rangle + \langle B_z\rangle \partial_z \langle v_\phi \rangle \right)}_{\text{stretching}} \\
    & + \underbrace{\partial_z \delta\mathcal{E}_R
    - \partial_R \delta\mathcal{E}_z}_{\text{turbulence}},
    \label{eq:dynamo_phi}
\end{aligned}
\end{equation}
including contributions from radial advection, vertical advection, stretching, and turbulence. 

In the bottom panels of \figu\ref{fig:dynamo}, we analyze the time and azimuthally averaged contribution of each component in \eq\ref{eq:dynamo_phi} for toroidal field $\langle B_\phi\rangle$ at $R/r_0=10$ and $R/r_0=30$ over the quasi-steady state ($3\times10^4 <t/t_0 < 6\times10^4$, that is, from 150 to 300 orbits at $R/r_0=10$). 
We normalize each term by $\langle B_\phi \rangle\Omega(R)/(2\pi)$ such that a value of $1$ implies that the particular term would grow or decay $\langle B_\phi\rangle$ at a given $z$ by 100\% over one orbit. 
For comparison, we show the evolution of the vertical profile of $B_\phi$ and $\sqrt{2P_\mathrm{gas}}$ at $R/r_0=10$ and $R/r_0=30$ in the top panels of \figu\ref{fig:dynamo}. The azimuthal field $B_\phi$ at $R=10\,r_0$ gradually increases, flips its sign after $\sim 100$ orbits, and saturates with $B_\phi\sim -\sqrt{2P_\mathrm{gas}}$. The system is in a statistically steady state, though the flip of sign at $R=10\,r_0$ complicates interpreting the time-averaged dynamo contributions to the evolution of the field. As we noted previously, the disk at $R=30\,r_0$ is still radially supported by rotation but is vertically supported by magnetic pressure with a much smaller $\beta\ll1$ and $\alpha \gg 1$ during the simulation. 

As shown in \figu\ref{fig:dynamo} (bottom left), at $R=10\,r_0$, the azimuthal magnetic field is continuously regenerated and escaping the disk. Similar to the local simulations \citep{Squire2024rapidstronglymagnetizedaccretion}, fields are replenished and expelled over several orbit timescales, with strong flux creation by shearing of radial field (dominated by the term $\langle B_R\rangle \partial_R \langle v_\phi\rangle$) balanced by its diffusion via turbulence (dominated by $\partial_z\delta\mathcal{E}_R$). The vertical advection of the azimuthal field is subdominant with its two subterms $-\langle B_\phi\rangle \partial_z \langle v_z \rangle / (\langle B_\phi\rangle \Omega/(2\pi)) \sim 0.3$ and $-\langle v_z\rangle \partial_z \langle B_\phi\rangle / (\langle B_\phi\rangle \Omega/(2\pi)) \sim -0.3$ canceling each other out. 
The radial advection of the azimuthal field is approximately zero, making it approximately equivalent to the shearing box. Though not shown in the figure, the first two subterms $-\langle v_\phi\rangle\langle B_R \rangle/R  / (\langle B_\phi\rangle \Omega/(2\pi))\sim 0.5 $ and  $ -\langle B_\phi \rangle \partial_R \langle v_R\rangle  / (\langle B_\phi\rangle \Omega/(2\pi)) \sim -0.5 $ are relatively large but cancel out each other while the third term $ -\langle v_R \rangle \partial_R \langle B_\phi\rangle / (\langle B_\phi\rangle \Omega/(2\pi)) \sim 0 $ is negligible. 

\figu\ref{fig:dynamo} (bottom right) shows the same analysis for the disk at $R=30\,r_0>R_\mathrm{circ}$. As in our analysis at $R = 10\,r_0$, the fields are still replenished by shearing of the radial magnetic field and expelled by turbulent diffusion. Nonetheless, the radial advection of the field $B_\phi$ (dominated by the term $-\langle B_\phi\rangle \partial_R \langle v_R\rangle$) replenishes the field quickly due to the relatively large radial velocity $-v_\mathrm{R}/v_\mathrm{K}\sim 0.1$. This may explain why the disk is able to maintain a $\beta\ll1$ state at larger radii. The fact that radial advection is only important exterior to the circularization radius implies that, at least in the simulations carried out here, radial advection of the magnetic field is important in setting the magnetic field structure at large radii but not in the bulk of the disk.

For the mean radial field $\langle B_R\rangle$, we have shown in \figu\ref{fig:Rz} that it is generated from the large-scale inflow of the gas. The large-scale inflow forms large-scale poloidal loops from the initially toroidal magnetic field, which leads to both mean radial and mean vertical magnetic field on smaller scales. Note, though, that the system still has zero net vertical flux since the mean vertical field is symmetric about the midplane.

\subsection{Robustness of the $\beta\sim1$ disk} \label{subsec:miscellaneous}

To investigate the robustness of the $\beta\sim1$ disk and its dependence on various factors, including the strength of the magnetic field, vertical magnetic field, and numerical method, we perform miscellaneous tests here.

\paragraph{Strength of the initial magnetic field} First, we vary the strength of the initial magnetic field to examine the robustness of the $\beta\sim1$ disk. We run a case (model h05b10) similar to the fiducial run except with a stronger initial magnetic field (lower $\beta=1/10$). The space-time diagram of midplane $\beta$ is shown in \figu\ref{fig:beta_misc}. Despite a ten times stronger initial magnetic field, the system still collapses to a $\beta\sim1$ disk within $\sim 80$ orbits, a timescale similar to the fiducial run. Therefore the $\beta\sim1$ state does not depend on the initial strength of the large-scale magnetic field as long as the initial magnetic field is strong enough to form the $\beta\sim1$ state during the circularization of the disk.

\paragraph{Net vertical magnetic field} As found by many previous studies~\citep{Hawley1995ApJ...440..742H,Bai2013ApJ...767...30B, Salvesen2016MNRAS.460.3488S.poloidal, Zhu2018ApJ...857...34Z}, including a net vertical magnetic field can help sustain a strongly magnetized ($\beta<1$) disk midplane. Here we perform a simulation with a purely vertical magnetic field with strength $\beta_z=P_\mathrm{gas}/(B_z^2/2)=2$ (model h05b0z1). The results are shown in \figu\ref{fig:beta_misc}. Consistent with previous studies, the system reaches a $\beta\lesssim1$ state with a larger density scale height. We also see the polarity flips of $B_\phi$ near the midplane and a similar structure of $v_R$ and $B_R$, as shown in \figu\ref{fig:b0z1_vertical}. The stress is dominated by the mean component of the Maxwell stress, $\alpha_{\langle B\rangle}$. Therefore, it is a rather different mode of accretion compared with the ZNVF cases, but consistent with the large-scale channel mode with vertical flux~\citep[e.g.,][]{Zhu2018ApJ...857...34Z}.
We also run a case (model h05b1z1) with both the toroidal and vertical field for completeness and show the results in \figu\ref{fig:beta_misc}. 
The final magnetization of the disk is qualitatively similar to the case with purely vertical fields. Therefore, similar to previous studies, the accretion with vertical field shows qualitatively different accretion properties from ZNVF cases. 

\paragraph{Reconstruction method} Most of the runs are performed using the PLM reconstruction method. Here we rerun the fiducial case but with the piecewise parabolic (PPM) reconstruction method (h05ppm) for $t/t_0=10^4$ to test the robustness of the results against the numerical method. The space-time diagram shown in \figu\ref{fig:beta_misc} shows a similar transition from $\beta\ll1$ to $\beta\sim1$ state, indicating that the results are very similar to the fiducial run.

\paragraph{Mesh refinement and derefinement} To further confirm that the transition of magnetic state is due to numerical resolution, we refine the mesh of the model h05n64 at $t/t_0=1.5\times10^4$ by a factor of 4, and run it for $5\times10^3\,t_0$ (model h05n64-256). Then we derefine the mesh back to the original resolution and run it for $10^4\,t_0$ (model h05n64-256-64). As shown in \figu\ref{fig:beta_misc}, the disk gradually collapses after several to tens of orbits with midplane plasma-$\beta$ gradually increasing to $\beta\sim1$ after refinement and returning to $\beta\ll1$ after derefinement. Therefore the $\beta\sim1$ state is clearly dependent on resolution. We do not find such a transition if we only refine the mesh of the model h05n64 by a factor of 2, implying that the dependence of the transition on resolution may not be very sharp. We also restart the fiducial run at $4000\,t_0$, refine it by a factor of 2, and run it for $3000\,t_0$ (model h05n512t0) to capture the initial collapse process in the fiducial run with a higher resolution. As shown in \figu\ref{fig:beta_misc}, the collapse process of the disk is very similar to the fiducial case.

\paragraph{Convergence of the solution} We assess convergence by comparing simulations with different resolutions relative to the thermal scale height and by restarting selected runs with refined or derefined meshes. These tests show that the qualitative disk state is primarily controlled by whether $H_\mathrm{th}$ is resolved: refining an initially under-resolved disk drives the transition to the $\beta\sim1$ state, while derefining it returns the disk to a $\beta\ll1$ state. For the fiducial run, we double the resolution both after the disk has reached the quasi-steady state (model h05n512t3) and during the initial collapse phase (model h05n512t0); in both cases the higher-resolution evolution is qualitatively similar to the fiducial run, as shown in \figu\ref{fig:beta_misc}. We therefore regard the existence of the resolved $\beta\sim1$, $\alpha\sim0.1$ state as numerically robust, although we do not claim strict convergence of quantities such as the saturated midplane $\beta$ and $\alpha$.

\paragraph{Initial turbulence} We also test whether the transition depends sensitively on the initially smooth inflow by adding initial large-scale velocity turbulence in model h10n128turb. As shown in \figu\ref{fig:beta_misc}, the disk still undergoes the collapse to a $\beta\sim1$ state, although on a somewhat longer timescale. This suggests that the magnetic transition is not simply an artifact of the initially smooth configuration, and that initial turbulence alone does not sustain a long-lived $\beta\ll1$ midplane state in this idealized setup.  However, a more detailed study of a range of turbulent initial conditions would be useful given that we have only explored one such initial condition here.

To summarize: The $\beta\sim 1$ disk is robust against various parameters. Varying thermal scale height, initial toroidal magnetic field strength, and numerical method give qualitatively similar $\beta\sim1$ disks within $\sim R_\mathrm{circ}$. A vertical field helps to sustain a different disk state with a larger saturated magnetic flux characterized by $\beta\lesssim1$. Further tests with smaller $H_\mathrm{th}$ would be particularly useful, but would require significantly higher resolution.

\section{Summary and Discussion}\label{sec:discussion}

We have presented and characterized three-dimensional idealized global MHD simulations of a strongly magnetized accretion disk dominated by toroidal magnetic fields. Our equation of state fixes the temperature via a specified value of the disk aspect ratio (i.e., $H/R$ assuming gas pressure support). 
The disk forms by gas inflow and circularization from uniform cold gas with a toroidal magnetic field. We find that the system maintains a strong mean azimuthal field in the midplane, with $\beta\sim1$, trans-Alfv\'enic fluctuations and large accretion stresses $\alpha\sim0.1$. The disk is turbulent with $\delta v^2\sim \delta v_A^2 \sim v_A^2 \sim c_s^2$, likely driven by a combination of MRI and Parker instabilities ~\citep{Balbus1992ApJ...400..610B, Kim2000ApJ...540..372K, Pessah2005ApJ...628..879P,Johansen2008AA...490..501J,Das2018MNRAS.473.2791D,Squire2024rapidstronglymagnetizedaccretion}. Outside the midplane, the disk is strongly magnetized with $\beta\ll1$.
A net vertical flux, which is traditionally thought to be necessary to sustain such strong magnetization ($\beta \sim 1$ in the midplane) and large $\alpha$, is not in fact required. That being said, we find that the addition of vertical magnetic flux can generate even stronger magnetization, as illustrated by previous works~\citep{Bai2013ApJ...767...30B, Salvesen2016MNRAS.460.3488S.poloidal, Zhu2018ApJ...857...34Z}.

We show explicitly that the azimuthal field in the disk is continuously escaping along the vertical direction but is also replenished via a local dynamo. 
In the bulk of the disk interior to the circularization radius, radial advection of azimuthal magnetic field from larger radii is subdominant in sustaining the $\beta \sim 1$ state and the toroidal field (though it is important exterior to the circularization radius where the gas inflows more rapidly). In this sense shearing box simulations should be a reasonable approximation to the disk dynamics found here. Indeed our global results are broadly similar to the ``low-$\beta$ state'' zero-net-vertical-flux shearing box results of \citet{Squire2024rapidstronglymagnetizedaccretion}.

We find that the gas first forms a highly magnetized ($\beta\ll1$) Keplerian disk, but then gradually collapses to a moderately magnetized ($\beta\sim1$) state from the inside out over $\sim 50-80$ orbits (\figu\ref{fig:pressure}).
The disk also can be strongly magnetized with $\beta\ll 1$ and $\alpha \gg 1$ when the thermal scale height is not resolved (\figus\ref{fig:all_beta_slice}, \ref{fig:all_beta}, \ref{fig:beta_slice}, and \ref{fig:alpha}). However, when we resolve the thermal scale height with $\gtrsim 4$ cells, the disk slowly collapses to a disk with $\beta\sim 1$ in the midplane over $\sim 50-80$ orbits at the corresponding radius; the disk remains magnetically dominated with $\beta \ll 1$ well off the midplane.
In the resolved runs, the dense midplane is close to thermal support, while much of the accretion occurs in the more strongly magnetized layers at $|z|\gtrsim H_\mathrm{th}$.

The disk properties presented here bear some similarities to the strongly magnetized accretion disks found in related global simulations by \citet{Gaburov2012ApJ...758..103G}, \citet{Hopkins2024OJAp....7E..18H, Hopkins2024OJAp....7E..19H} and \citet{Guo2024arXiv240511711G}. \citet{Gaburov2012ApJ...758..103G} found the formation of a toroidally dominated $\beta\sim 0.1$ disk in simulations of accretion disks formed by molecular cloud disruption in galactic nuclei. \citet{Hopkins2024OJAp....7E..18H, Hopkins2024OJAp....7E..19H} performed simulations following the accretion flow onto a supermassive black hole (SMBH) in cosmological simulations from $\sim\,\mathrm{Mpc}$ scales and found a disk that is highly magnetized ($\beta\sim10^{-4}$), toroidal field-dominated, vertically supported by the strong magnetic pressure, with $\alpha\gg1$ due to the strong magnetic stress. \citet{Hopkins2025OJAp....8E..48H} re-ran a new case with a higher temperature floor of $10^6\,\mathrm{K}$ (see their Fig. 31) so that thermal scale height is resolved; they still found a strongly magnetized state with $\beta\ll1$ after $\sim 50$ orbits at inner boundary. \citet{Guo2024arXiv240511711G} found similar strongly-magnetized disks with $\beta\sim 10^{-3}$ when following the fueling of SMBH in elliptical galaxies from a turbulent cooling medium on galactic scales. 

Overall, our idealized simulations capture a number of these features, such as the dominant toroidal field and trans-Alfv\'enic fluctuations. Most importantly, our results show that the magnetic flux does not immediately escape vertically, as previously argued from both local and global simulations~\citep{Salvesen2016MNRAS.460.3488S.poloidal, Fragile2017MNRAS.467.1838F}; rather, the disk can sustain a  $\beta\sim1$ field against rapid escape via a dynamo mechanism. However, a critical difference is that the midplane $\beta$ in our simulations is markedly higher than the solutions in \citet{Hopkins2024OJAp....7E..19H} and \citet{Guo2024arXiv240511711G}, with $\beta \sim 1$ instead of $\beta\ll1$. As in the local shearing box simulations of \citet{Squire2024rapidstronglymagnetizedaccretion}, the only situation in which we sustain $\beta \ll 1$ in the bulk of the disk for the full duration of our global simulations is when the midplane thermal scale height of the gas is poorly resolved (\figu\ref{fig:pressure}). The similarity of our global simulations to shearing box simulations in this respect shows that radial advection of magnetic flux does not appear to generically maintain a $\beta \ll 1$ toroidal field given such a state at larger radii. 

The difference between our global simulations and those of \citet{Hopkins2024OJAp....7E..19H} and \citet{Guo2024arXiv240511711G} remains unclear. It may be that the latter are simply not well enough resolved -- this could reconcile the differences -- or it could be that the more complex gas thermodynamics and gas inflow properties including strong asymmetry, warps, and more turbulence in the more realistic global models change the efficiency of maintaining the $\beta \ll 1$ state. It is striking that our global simulations maintain $\beta \ll 1$ and $\alpha \gg 1$ for dozens of dynamical times (orbits) in the earlier phase when the accretion rate is increasing rapidly in time so that the ambient disk conditions are evolving on the timescale that magnetic flux is escaping (\figus~\ref{fig:hist} and \ref{fig:mdot}). The simulations also maintain $\alpha\gtrsim 1$ somewhat exterior to the circularization radius (but still in the rotationally dominated part of the flow). These may be clues regarding conditions under which disks can maintain $\beta \ll 1$. \citet{Tomar2026OJAp....960174T} recently repeated this idealized toroidally magnetized disk problem with Lagrangian methods and found that, when the thermal scale height is resolved, their simulations also evolve toward the flux-losing, $\beta\sim1$ state found here. This suggests that the transition is not simply a consequence of using a static Eulerian mesh. More work on understanding the connection of these idealized problems to the related global simulations of \citet{Hopkins2024OJAp....7E..19H, Guo2024arXiv240511711G} is clearly needed.

The strong toroidal magnetic field state found in \citet{Squire2024rapidstronglymagnetizedaccretion} and here may naturally explain the efficiency of angular momentum transport $\alpha\sim0.1-0.4$  estimated in dwarf novae and X-ray transient outbursts \citep{King2007MNRAS.376.1740K, Tetarenko2018Natur.554...69T}.
While the inclusion of other physics can do so as well (e.g., \citealt{Hirose2014ApJ...787....1H, Begelman2015ApJ...809..118B, Scepi2018A&A...620A..49S}), the strong toroidal field disks studied in \citet{Squire2024rapidstronglymagnetizedaccretion} and here appear to produce $\alpha \sim 0.1-0.5$ relatively independent of other aspects of the simulation.

It is not clear if the results presented here are fully converged. Local simulations in \citet{Squire2024rapidstronglymagnetizedaccretion} suggest that the midplane $\beta$ and $\alpha$ are non-monotonic with resolution (see their Fig. 17). With resolution increasing, the disk transport level first suddenly decreases from $\alpha\sim1$ to $\alpha \sim0.1$ then gradually increases with increasing resolution to $\alpha \simeq 0.5$ at the highest resolution studied.
A resolution of at least 10 cells per thermal scale height is required to approach convergence in the local shearing box simulations; this is a resolution $\sim2$ times better than the fiducial case here in which case it is possible that we actually underestimate the saturated magnetic field strength and $\alpha$ by a modest amount. Our global simulations also have limited dynamic range between the circularization radius and inner radius, and between the disk scale height and local disk radius.
Further simulations with larger separation between $R_\mathrm{circ}$ and $r_\mathrm{in}$, smaller thermal scale height, and higher resolution would be valuable.

The simulations in this work are performed in an idealized Newtonian framework to focus on the effects of strong toroidal magnetic fields while enabling high-resolution, long-duration global calculations. This setup is most applicable to disk regions where relativistic effects are subdominant. In the inner regions of realistic accretion disks, general relativistic effects such as the innermost stable circular orbit (ISCO), plunging inflow, and (for spinning black holes) frame dragging may modify magnetic flux transport and introduce additional phenomena such as warping or disk tearing, which are not included here. We also employ a sink region as the inner boundary, where the sink radius is not intended to represent the true event horizon, and thus boundary effects may influence the very inner flow. For the inner regions of black hole accretion systems, GRMHD simulations~\citep{Fragile2017MNRAS.467.1838F,Rule2025MNRAS.542..377R} would therefore be valuable.

We note that additional physical processes not included in this study, such as radiative transport and dissipative physics, may modify the quantitative conditions for the existence and stability of strongly magnetized disks by altering the local thermal balance. In particular, these effects can shift the precise instability threshold but are not expected to change the qualitative role of strong toroidal magnetic fields in the disk dynamics. Self-gravity is neglected in this work for simplicity but may become important in the outer disk at sufficiently high accretion rates. These aspects are beyond the scope of the present idealized study and are left for future work.

\begin{acknowledgments}
We thank Yuri Levin for useful conversations. We thank the anonymous referee for the helpful comments and suggestions. MG thanks Yixian Chen for helpful discussions. We acknowledge the EuroHPC Joint Undertaking for awarding this project access to the EuroHPC supercomputer LUMI, hosted by CSC (Finland) and the LUMI consortium through a EuroHPC Regular Access call. The authors are pleased to acknowledge that the work reported on in this paper was substantially performed using the Princeton Research Computing resources at Princeton University, which is consortium of groups led by the Princeton Institute for Computational Science and Engineering (PICSciE) and Office of Information Technology's Research Computing. This work was supported in part by NSF AST grant 2107872, by a Simons Investigator grant to EQ, and by a grant from the Simons Foundation (888968, E.C. Ostriker, Princeton University PI) as part of the Learning the Universe Collaboration. JS acknowledges the support of the Royal Society Te Ap\=arangi, through Marsden-Fund grant MFP-UOO2221 and Rutherford Discovery Fellowship RDF-U001804. This work benefited from EQ's and JS's stays at the Kavli Institute for Theoretical Physics, supported by NSF PHY-2309135.
\end{acknowledgments}

\vspace{5mm}
{Software: \athenak{} \citep{Stone2026ApJS..283...27S}}

\appendix

\section{Resolution of the simulations} \label{app:resolution}

\begin{figure}[ht]
\centering
\includegraphics[width=\linewidth]{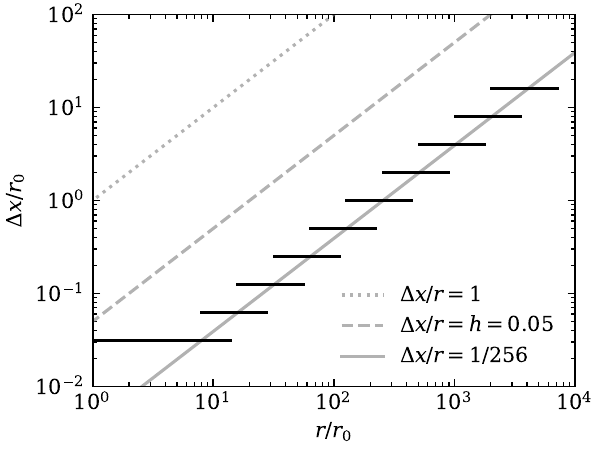}
\caption{Size of cells (black) as a function of spherical radius in the highest resolution run. The overlap of the black lines is from the conversion from Cartesian to spherical coordinates. Gray lines mark characteristic resolution of $\Delta x/r=1$, $0.05$, and $1/256$. The resolution for other runs is adjusted everywhere accordingly.
\label{fig:resolution}}
\end{figure}

For clarity, \figu\ref{fig:resolution} illustrates the resolution of the highest resolution runs (models h05n512t3 and h05n512t0). The resolution for other simulations is changed everywhere accordingly. The resolution is roughly uniform logarithmically. The highest resolution covers a region of $r\lesssim 10\,r_0$.

\end{CJK*}

\bibliographystyle{mn2e}
\bibliography{main}

@ARTICLE{Tomar2026OJAp....960174T,
       author = {{Tomar}, Yashvardhan and {Hopkins}, Philip F.},
        title = "{Lagrangian versus Eulerian Methods for Toroidally-Magnetized Isothermal Disks}",
      journal = {The Open Journal of Astrophysics},
         year = 2026,
        month = apr,
       volume = {9},
        pages = {60174},
          doi = {10.33232/001c.160174},
archivePrefix = {arXiv},
       eprint = {2512.05194},
 primaryClass = {astro-ph.HE},
       adsurl = {https://ui.adsabs.harvard.edu/abs/2026OJAp....960174T}
}

@ARTICLE{Rule2025MNRAS.542..377R,
       author = {{Rule}, Jake and {Mummery}, Andrew and {Balbus}, Steven and {Stone}, James M. and {Zhang}, Lizhong},
        title = "{The plunging region of a thin accretion disc around a Schwarzschild black hole}",
      journal = {\mnras},
         year = 2025,
        month = sep,
       volume = {542},
       number = {1},
        pages = {377-390},
          doi = {10.1093/mnras/staf1256},
archivePrefix = {arXiv},
       eprint = {2505.13701},
 primaryClass = {astro-ph.HE},
       adsurl = {https://ui.adsabs.harvard.edu/abs/2025MNRAS.542..377R}
}

@ARTICLE{Jiang2025arXiv250509671J,
       author = {{Jiang}, Yan-Fei and {Blaes}, Omer and {Kaul}, Ish and {Zhang}, Lizhong},
        title = "{Radiation and Magnetic Pressure Support in Accretion Disks around Supermassive Black Holes and The Physical Origin of the Extreme Ultraviolet to Soft X-ray Spectrum}",
      journal = {arXiv e-prints},
         year = 2025,
        month = may,
          eid = {arXiv:2505.09671},
        pages = {arXiv:2505.09671},
          doi = {10.48550/arXiv.2505.09671},
archivePrefix = {arXiv},
       eprint = {2505.09671},
 primaryClass = {astro-ph.HE},
       adsurl = {https://ui.adsabs.harvard.edu/abs/2025arXiv250509671J}
}

@ARTICLE{Wang2025arXiv250403874W,
       author = {{Wang}, Hai-Yang and {Guo}, Minghao and {Most}, Elias R. and {Hopkins}, Philip F. and {Lalakos}, Aretaios},
        title = "{Galactic-scale Feeding Reveals Warped Hypermagnetized Multiphase Circumbinary Accretion Around Supermassive Black Hole Binaries}",
      journal = {arXiv e-prints},
         year = 2025,
        month = apr,
          eid = {arXiv:2504.03874},
        pages = {arXiv:2504.03874},
          doi = {10.48550/arXiv.2504.03874},
archivePrefix = {arXiv},
       eprint = {2504.03874},
 primaryClass = {astro-ph.HE},
       adsurl = {https://ui.adsabs.harvard.edu/abs/2025arXiv250403874W}
}

@ARTICLE{Hopkins2025OJAp....8E..48H,
       author = {{Hopkins}, Philip F. and {Su}, Kung-Yi and {Murray}, Norman and {Steinwandel}, Ulrich P. and {Kaaz}, Nicholas and {Ponnada}, Sam B. and {Bardati}, Jaeden and {Piotrowska}, Joanna M. and {Wang}, Hai-Yang and {Shi}, Yanlong and {Angles-Alcazar}, Daniel and {Most}, Elias R. and {Kremer}, Kyle and {Faucher-Giguere}, Claude-Andre and {Wellons}, Sarah},
        title = "{Zooming In On The Multi-Phase Structure of Magnetically-Dominated Quasar Disks: Radiation From Torus to ISCO Across Accretion Rates}",
      journal = {The Open Journal of Astrophysics},
         year = 2025,
        month = apr,
       volume = {8},
          eid = {48},
        pages = {48},
          doi = {10.33232/001c.137296},
archivePrefix = {arXiv},
       eprint = {2502.05268},
 primaryClass = {astro-ph.GA},
       adsurl = {https://ui.adsabs.harvard.edu/abs/2025OJAp....8E..48H}
}

@ARTICLE{Ryan2017ApJ...840....6R,
       author = {{Ryan}, Benjamin R. and {Gammie}, Charles F. and {Fromang}, Sebastien and {Kestener}, Pierre},
        title = "{Resolution Dependence of Magnetorotational Turbulence in the Isothermal Stratified Shearing Box}",
      journal = {\apj},
         year = 2017,
        month = may,
       volume = {840},
       number = {1},
          eid = {6},
        pages = {6},
          doi = {10.3847/1538-4357/aa6a52},
archivePrefix = {arXiv},
       eprint = {1702.00777},
 primaryClass = {astro-ph.HE},
       adsurl = {https://ui.adsabs.harvard.edu/abs/2017ApJ...840....6R}
}

@ARTICLE{Sadowski2016MNRAS.459.4397S,
       author = {{S{\k{a}}dowski}, Aleksander},
        title = "{Thin accretion discs are stabilized by a strong magnetic field}",
      journal = {\mnras},
         year = 2016,
        month = jul,
       volume = {459},
       number = {4},
        pages = {4397-4407},
          doi = {10.1093/mnras/stw913},
archivePrefix = {arXiv},
       eprint = {1601.06785},
 primaryClass = {astro-ph.HE},
       adsurl = {https://ui.adsabs.harvard.edu/abs/2016MNRAS.459.4397S}
}

@ARTICLE{Machida2006PASJ...58..193M,
       author = {{Machida}, Mami and {Nakamura}, Kenji E. and {Matsumoto}, Ryoji},
        title = "{Formation of Magnetically Supported Disks during Hard-to-Soft Transitions in Black Hole Accretion Flows}",
      journal = {\pasj},
         year = 2006,
        month = feb,
       volume = {58},
        pages = {193-202},
          doi = {10.1093/pasj/58.1.193},
archivePrefix = {arXiv},
       eprint = {astro-ph/0511299},
 primaryClass = {astro-ph},
       adsurl = {https://ui.adsabs.harvard.edu/abs/2006PASJ...58..193M}
}

@BOOK{Frank2002apa..book.....F,
       author = {{Frank}, Juhan and {King}, Andrew and {Raine}, Derek J.},
        title = "{Accretion Power in Astrophysics: Third Edition}",
         year = 2002,
       adsurl = {https://ui.adsabs.harvard.edu/abs/2002apa..book.....F}
}

@ARTICLE{Goodman2003,
       author = {{Goodman}, Jeremy},
        title = "{Self-gravity and quasi-stellar object discs}",
      journal = {\mnras},
         year = 2003,
        month = mar,
       volume = {339},
       number = {4},
        pages = {937-948},
          doi = {10.1046/j.1365-8711.2003.06241.x},
archivePrefix = {arXiv},
       eprint = {astro-ph/0201001},
 primaryClass = {astro-ph},
       adsurl = {https://ui.adsabs.harvard.edu/abs/2003MNRAS.339..937G}
}

@ARTICLE{Shlosman1987Natur.329..810S,
       author = {{Shlosman}, Isaac and {Begelman}, Mitchell C.},
        title = "{Self-gravitating accretion disks in active galactic nuclei}",
      journal = {\nat},
         year = 1987,
        month = oct,
       volume = {329},
       number = {6142},
        pages = {810-812},
          doi = {10.1038/329810a0},
       adsurl = {https://ui.adsabs.harvard.edu/abs/1987Natur.329..810S}
}

@ARTICLE{Parker1958ApJ...128..664P,
       author = {{Parker}, E.~N.},
        title = "{Dynamics of the Interplanetary Gas and Magnetic Fields.}",
      journal = {\apj},
         year = 1958,
        month = nov,
       volume = {128},
        pages = {664},
          doi = {10.1086/146579},
       adsurl = {https://ui.adsabs.harvard.edu/abs/1958ApJ...128..664P}
}

@misc{Squire2024rapidstronglymagnetizedaccretion,
      title={Rapid, strongly magnetized accretion in the zero-net-vertical-flux shearing box}, 
      author={Jonathan Squire and Eliot Quataert and Philip F. Hopkins},
      year={2024},
      eprint={2409.05467},
      archivePrefix={arXiv},
      primaryClass={astro-ph.HE},
      url={https://arxiv.org/abs/2409.05467}, 
}

@ARTICLE{Scepi2018A&A...620A..49S,
       author = {{Scepi}, N. and {Lesur}, G. and {Dubus}, G. and {Flock}, M.},
        title = "{Turbulent and wind-driven accretion in dwarf novae threaded by a large-scale magnetic field}",
      journal = {\aap},
         year = 2018,
        month = dec,
       volume = {620},
          eid = {A49},
        pages = {A49},
          doi = {10.1051/0004-6361/201833921},
archivePrefix = {arXiv},
       eprint = {1809.09131},
 primaryClass = {astro-ph.HE},
       adsurl = {https://ui.adsabs.harvard.edu/abs/2018A&A...620A..49S}
}

@ARTICLE{Begelman2015ApJ...809..118B,
       author = {{Begelman}, Mitchell C. and {Armitage}, Philip J. and {Reynolds}, Christopher S.},
        title = "{Accretion Disk Dynamo as the Trigger for X-Ray Binary State Transitions}",
      journal = {\apj},
         year = 2015,
        month = aug,
       volume = {809},
       number = {2},
          eid = {118},
        pages = {118},
          doi = {10.1088/0004-637X/809/2/118},
archivePrefix = {arXiv},
       eprint = {1507.03996},
 primaryClass = {astro-ph.HE},
       adsurl = {https://ui.adsabs.harvard.edu/abs/2015ApJ...809..118B}
}

@ARTICLE{Hirose2014ApJ...787....1H,
       author = {{Hirose}, Shigenobu and {Blaes}, Omer and {Krolik}, Julian H. and {Coleman}, Matthew S.~B. and {Sano}, Takayoshi},
        title = "{Convection Causes Enhanced Magnetic Turbulence in Accretion Disks in Outburst}",
      journal = {\apj},
         year = 2014,
        month = may,
       volume = {787},
       number = {1},
          eid = {1},
        pages = {1},
          doi = {10.1088/0004-637X/787/1/1},
archivePrefix = {arXiv},
       eprint = {1403.3096},
 primaryClass = {astro-ph.HE},
       adsurl = {https://ui.adsabs.harvard.edu/abs/2014ApJ...787....1H}
}

@ARTICLE{Kudoh2020ApJ...904....9K,
       author = {{Kudoh}, Yuki and {Wada}, Keiichi and {Norman}, Colin},
        title = "{Multiphase Circumnuclear Gas in a Low-{\ensuremath{\beta}} Disk: Turbulence and Magnetic Field Reversals}",
      journal = {\apj},
         year = 2020,
        month = nov,
       volume = {904},
       number = {1},
          eid = {9},
        pages = {9},
          doi = {10.3847/1538-4357/abba39},
archivePrefix = {arXiv},
       eprint = {2008.07050},
 primaryClass = {astro-ph.GA},
       adsurl = {https://ui.adsabs.harvard.edu/abs/2020ApJ...904....9K}
}

@ARTICLE{Fragile2017MNRAS.467.1838F,
       author = {{Fragile}, P. Chris and {S{\k{a}}dowski}, Aleksander},
        title = "{On the decay of strong magnetization in global disc simulations with toroidal fields}",
      journal = {\mnras},
         year = 2017,
        month = may,
       volume = {467},
       number = {2},
        pages = {1838-1843},
          doi = {10.1093/mnras/stx274},
archivePrefix = {arXiv},
       eprint = {1701.01159},
 primaryClass = {astro-ph.HE},
       adsurl = {https://ui.adsabs.harvard.edu/abs/2017MNRAS.467.1838F}
}

@ARTICLE{Blandford1982MNRAS.199..883B,
       author = {{Blandford}, R.~D. and {Payne}, D.~G.},
        title = "{Hydromagnetic flows from accretion disks and the production of radio jets.}",
      journal = {\mnras},
         year = 1982,
        month = jun,
       volume = {199},
        pages = {883-903},
          doi = {10.1093/mnras/199.4.883},
       adsurl = {https://ui.adsabs.harvard.edu/abs/1982MNRAS.199..883B}
}

@ARTICLE{Zhu2018ApJ...857...34Z,
       author = {{Zhu}, Zhaohuan and {Stone}, James M.},
        title = "{Global Evolution of an Accretion Disk with a Net Vertical Field: Coronal Accretion, Flux Transport, and Disk Winds}",
      journal = {\apj},
         year = 2018,
        month = apr,
       volume = {857},
       number = {1},
          eid = {34},
        pages = {34},
          doi = {10.3847/1538-4357/aaafc9},
archivePrefix = {arXiv},
       eprint = {1701.04627},
 primaryClass = {astro-ph.EP},
       adsurl = {https://ui.adsabs.harvard.edu/abs/2018ApJ...857...34Z}
}

@ARTICLE{Lesur2013A&A...550A..61L,
       author = {{Lesur}, G. and {Ferreira}, J. and {Ogilvie}, G.~I.},
        title = "{The magnetorotational instability as a jet launching mechanism}",
      journal = {\aap},
         year = 2013,
        month = feb,
       volume = {550},
          eid = {A61},
        pages = {A61},
          doi = {10.1051/0004-6361/201220395},
archivePrefix = {arXiv},
       eprint = {1210.6660},
 primaryClass = {astro-ph.HE},
       adsurl = {https://ui.adsabs.harvard.edu/abs/2013A&A...550A..61L}
}

@ARTICLE{Gaburov2012ApJ...758..103G,
       author = {{Gaburov}, Evghenii and {Johansen}, Anders and {Levin}, Yuri},
        title = "{Magnetically Levitating Accretion Disks around Supermassive Black Holes}",
      journal = {\apj},
         year = 2012,
        month = oct,
       volume = {758},
       number = {2},
          eid = {103},
        pages = {103},
          doi = {10.1088/0004-637X/758/2/103},
archivePrefix = {arXiv},
       eprint = {1201.4873},
 primaryClass = {astro-ph.GA},
       adsurl = {https://ui.adsabs.harvard.edu/abs/2012ApJ...758..103G}
}

@ARTICLE{Begelman2007MNRAS.375.1070B,
       author = {{Begelman}, M.~C. and {Pringle}, J.~E.},
        title = "{Accretion discs with strong toroidal magnetic fields}",
      journal = {\mnras},
         year = 2007,
        month = mar,
       volume = {375},
       number = {3},
        pages = {1070-1076},
          doi = {10.1111/j.1365-2966.2006.11372.x},
archivePrefix = {arXiv},
       eprint = {astro-ph/0612300},
 primaryClass = {astro-ph},
       adsurl = {https://ui.adsabs.harvard.edu/abs/2007MNRAS.375.1070B}
}

@ARTICLE{Tetarenko2018Natur.554...69T,
       author = {{Tetarenko}, B.~E. and {Lasota}, J. -P. and {Heinke}, C.~O. and {Dubus}, G. and {Sivakoff}, G.~R.},
        title = "{Strong disk winds traced throughout outbursts in black-hole X-ray binaries}",
      journal = {\nat},
         year = 2018,
        month = feb,
       volume = {554},
       number = {7690},
        pages = {69-72},
          doi = {10.1038/nature25159},
archivePrefix = {arXiv},
       eprint = {1801.07203},
 primaryClass = {astro-ph.HE},
       adsurl = {https://ui.adsabs.harvard.edu/abs/2018Natur.554...69T}
}

@ARTICLE{King2007MNRAS.376.1740K,
       author = {{King}, A.~R. and {Pringle}, J.~E. and {Livio}, M.},
        title = "{Accretion disc viscosity: how big is alpha?}",
      journal = {\mnras},
         year = 2007,
        month = apr,
       volume = {376},
       number = {4},
        pages = {1740-1746},
          doi = {10.1111/j.1365-2966.2007.11556.x},
archivePrefix = {arXiv},
       eprint = {astro-ph/0701803},
 primaryClass = {astro-ph},
       adsurl = {https://ui.adsabs.harvard.edu/abs/2007MNRAS.376.1740K}
}

@ARTICLE{Bai2013ApJ...767...30B,
       author = {{Bai}, Xue-Ning and {Stone}, James M.},
        title = "{Local Study of Accretion Disks with a Strong Vertical Magnetic Field: Magnetorotational Instability and Disk Outflow}",
      journal = {\apj},
         year = 2013,
        month = apr,
       volume = {767},
       number = {1},
          eid = {30},
        pages = {30},
          doi = {10.1088/0004-637X/767/1/30},
archivePrefix = {arXiv},
       eprint = {1210.6661},
 primaryClass = {astro-ph.HE},
       adsurl = {https://ui.adsabs.harvard.edu/abs/2013ApJ...767...30B}
}

@ARTICLE{Hawley1995ApJ...440..742H,
       author = {{Hawley}, John F. and {Gammie}, Charles F. and {Balbus}, Steven A.},
        title = "{Local Three-dimensional Magnetohydrodynamic Simulations of Accretion Disks}",
      journal = {\apj},
         year = 1995,
        month = feb,
       volume = {440},
        pages = {742},
          doi = {10.1086/175311},
       adsurl = {https://ui.adsabs.harvard.edu/abs/1995ApJ...440..742H}
}

@ARTICLE{Stone1996ApJ...463..656S,
       author = {{Stone}, James M. and {Hawley}, John F. and {Gammie}, Charles F. and {Balbus}, Steven A.},
        title = "{Three-dimensional Magnetohydrodynamical Simulations of Vertically Stratified Accretion Disks}",
      journal = {\apj},
         year = 1996,
        month = jun,
       volume = {463},
        pages = {656},
          doi = {10.1086/177280},
       adsurl = {https://ui.adsabs.harvard.edu/abs/1996ApJ...463..656S}
}

@ARTICLE{Balbus1991ApJ...376..214B,
       author = {{Balbus}, Steven A. and {Hawley}, John F.},
        title = "{A Powerful Local Shear Instability in Weakly Magnetized Disks. I. Linear Analysis}",
      journal = {\apj},
         year = 1991,
        month = jul,
       volume = {376},
        pages = {214},
          doi = {10.1086/170270},
       adsurl = {https://ui.adsabs.harvard.edu/abs/1991ApJ...376..214B}
}

@ARTICLE{Shakura1973A&A....24..337S,
       author = {{Shakura}, N.~I. and {Sunyaev}, R.~A.},
        title = "{Black holes in binary systems. Observational appearance.}",
      journal = {\aap},
         year = 1973,
        month = jan,
       volume = {24},
        pages = {337-355},
       adsurl = {https://ui.adsabs.harvard.edu/abs/1973A&A....24..337S}
}

@ARTICLE{Guo2024arXiv240511711G,
       author = {{Guo}, Minghao and {Stone}, James M. and {Quataert}, Eliot and {Kim}, Chang-Goo},
        title = "{Magnetized Accretion onto and Feedback from Supermassive Black Holes in Elliptical Galaxies}",
      journal = {arXiv e-prints},
         year = 2024,
        month = may,
          eid = {arXiv:2405.11711},
        pages = {arXiv:2405.11711},
          doi = {10.48550/arXiv.2405.11711},
archivePrefix = {arXiv},
       eprint = {2405.11711},
 primaryClass = {astro-ph.HE},
       adsurl = {https://ui.adsabs.harvard.edu/abs/2024arXiv240511711G}
}

@ARTICLE{Mishra2020MNRAS.492.1855M,
       author = {{Mishra}, Bhupendra and {Begelman}, Mitchell C. and {Armitage}, Philip J. and {Simon}, Jacob B.},
        title = "{Strongly magnetized accretion discs: structure and accretion from global magnetohydrodynamic simulations}",
      journal = {\mnras},
         year = 2020,
        month = feb,
       volume = {492},
       number = {2},
        pages = {1855-1868},
          doi = {10.1093/mnras/stz3572},
archivePrefix = {arXiv},
       eprint = {1907.08995},
 primaryClass = {astro-ph.HE},
       adsurl = {https://ui.adsabs.harvard.edu/abs/2020MNRAS.492.1855M}
}

@ARTICLE{Das2018MNRAS.473.2791D,
       author = {{Das}, Upasana and {Begelman}, Mitchell C. and {Lesur}, Geoffroy},
        title = "{Instability in strongly magnetized accretion discs: a global perspective}",
      journal = {\mnras},
         year = 2018,
        month = jan,
       volume = {473},
       number = {2},
        pages = {2791-2812},
          doi = {10.1093/mnras/stx2518},
archivePrefix = {arXiv},
       eprint = {1709.09173},
 primaryClass = {astro-ph.HE},
       adsurl = {https://ui.adsabs.harvard.edu/abs/2018MNRAS.473.2791D}
}

@ARTICLE{Pessah2005ApJ...628..879P,
       author = {{Pessah}, Martin E. and {Psaltis}, Dimitrios},
        title = "{The Stability of Magnetized Rotating Plasmas with Superthermal Fields}",
      journal = {\apj},
         year = 2005,
        month = aug,
       volume = {628},
       number = {2},
        pages = {879-901},
          doi = {10.1086/430940},
archivePrefix = {arXiv},
       eprint = {astro-ph/0406071},
 primaryClass = {astro-ph},
       adsurl = {https://ui.adsabs.harvard.edu/abs/2005ApJ...628..879P}
}

@ARTICLE{Salvesen2016MNRAS.460.3488S.poloidal,
       author = {{Salvesen}, Greg and {Armitage}, Philip J. and {Simon}, Jacob B. and {Begelman}, Mitchell C.},
        title = "{Strongly magnetized accretion discs require poloidal flux}",
      journal = {\mnras},
         year = 2016,
        month = aug,
       volume = {460},
       number = {4},
        pages = {3488-3493},
          doi = {10.1093/mnras/stw1231},
archivePrefix = {arXiv},
       eprint = {1602.04810},
 primaryClass = {astro-ph.HE},
       adsurl = {https://ui.adsabs.harvard.edu/abs/2016MNRAS.460.3488S}
}

@ARTICLE{Johansen2008AA...490..501J,
       author = {{Johansen}, A. and {Levin}, Y.},
        title = "{High accretion rates in magnetised Keplerian discs mediated by a Parker instability driven dynamo}",
      journal = {\aap},
         year = 2008,
        month = nov,
       volume = {490},
       number = {2},
        pages = {501-514},
          doi = {10.1051/0004-6361:200810385},
archivePrefix = {arXiv},
       eprint = {0808.3579},
 primaryClass = {astro-ph},
       adsurl = {https://ui.adsabs.harvard.edu/abs/2008A&A...490..501J}
}

@ARTICLE{Hopkins2024OJAp....7E..19H,
       author = {{Hopkins}, Philip F. and {Squire}, Jonathan and {Su}, Kung-Yi and {Steinwandel}, Ulrich P. and {Kremer}, Kyle and {Shi}, Yanlong and {Grudic}, Michael Y. and {Wellons}, Sarah and {Faucher-Giguere}, Claude-Andre and {Angles-Alcazar}, Daniel and {Murray}, Norman and {Quataert}, Eliot},
        title = "{FORGE'd in FIRE II: The Formation of Magnetically-Dominated Quasar Accretion Disks from Cosmological Initial Conditions}",
      journal = {The Open Journal of Astrophysics},
         year = 2024,
        month = mar,
       volume = {7},
          eid = {19},
        pages = {19},
          doi = {10.21105/astro.2310.04506},
archivePrefix = {arXiv},
       eprint = {2310.04506},
 primaryClass = {astro-ph.HE},
       adsurl = {https://ui.adsabs.harvard.edu/abs/2024OJAp....7E..19H}
}

@ARTICLE{Hopkins2024OJAp....7E..18H,
       author = {{Hopkins}, Philip F. and {Grudic}, Michael Y. and {Su}, Kung-Yi and {Wellons}, Sarah and {Angles-Alcazar}, Daniel and {Steinwandel}, Ulrich P. and {Guszejnov}, David and {Murray}, Norman and {Faucher-Giguere}, Claude-Andre and {Quataert}, Eliot and {Keres}, Dusan},
        title = "{FORGE'd in FIRE: Resolving the End of Star Formation and Structure of AGN Accretion Disks from Cosmological Initial Conditions}",
      journal = {The Open Journal of Astrophysics},
         year = 2024,
        month = mar,
       volume = {7},
          eid = {18},
        pages = {18},
          doi = {10.21105/astro.2309.13115},
archivePrefix = {arXiv},
       eprint = {2309.13115},
 primaryClass = {astro-ph.GA},
       adsurl = {https://ui.adsabs.harvard.edu/abs/2024OJAp....7E..18H}
}

@ARTICLE{Kim2000ApJ...540..372K,
       author = {{Kim}, Woong-Tae and {Ostriker}, Eve C.},
        title = "{Magnetohydrodynamic Instabilities in Shearing, Rotating, Stratified Winds and Disks}",
      journal = {\apj},
         year = 2000,
        month = sep,
       volume = {540},
       number = {1},
        pages = {372-403},
          doi = {10.1086/309293},
archivePrefix = {arXiv},
       eprint = {astro-ph/0004094},
 primaryClass = {astro-ph},
       adsurl = {https://ui.adsabs.harvard.edu/abs/2000ApJ...540..372K}
}

@ARTICLE{Balbus1992ApJ...400..610B,
       author = {{Balbus}, Steven A. and {Hawley}, John F.},
        title = "{A Powerful Local Shear Instability in Weakly Magnetized Disks. IV. Nonaxisymmetric Perturbations}",
      journal = {\apj},
         year = 1992,
        month = dec,
       volume = {400},
        pages = {610-621},
          doi = {10.1086/172022},
       adsurl = {https://ui.adsabs.harvard.edu/abs/1992ApJ...400..610B}
}

@ARTICLE{Trott2021CSE....23e..10T,
       author = {{Trott}, Christian and {Berger-Vergiat}, Luc and {Poliakoff}, David and {Rajamanickam}, Sivasankaran and {Lebrun-Grandie}, Damien and {Madsen}, Jonathan and {Al Awar}, Nader and {Gligoric}, Milos and {Shipman}, Galen and {Womeldorff}, Geoff},
        title = "{The Kokkos EcoSystem: Comprehensive Performance Portability for High Performance Computing}",
      journal = {Computing in Science and Engineering},
         year = 2021,
        month = sep,
       volume = {23},
       number = {5},
        pages = {10-18},
          doi = {10.1109/MCSE.2021.3098509},
       adsurl = {https://ui.adsabs.harvard.edu/abs/2021CSE....23e..10T}
}

@ARTICLE{Stone2026ApJS..283...27S,
       author = {{Stone}, James M. and {Mullen}, Patrick D. and {Fielding}, Drummond and {Grete}, Philipp and {Guo}, Minghao and {Kempski}, Philipp and {Most}, Elias R. and {White}, Christopher J. and {Wong}, George N.},
        title = "{AthenaK: A Performance-portable Version of the Athena++ Adaptive Mesh Refinement Framework}",
      journal = {\apjs},
         year = 2026,
        month = mar,
       volume = {283},
       number = {1},
          eid = {27},
        pages = {27},
          doi = {10.3847/1538-4365/ae3717},
archivePrefix = {arXiv},
       eprint = {2409.16053},
 primaryClass = {astro-ph.IM},
       adsurl = {https://ui.adsabs.harvard.edu/abs/2026ApJS..283...27S}
}

@ARTICLE{Stone2020ApJS..249....4S,
       author = {{Stone}, James M. and {Tomida}, Kengo and {White}, Christopher J. and {Felker}, Kyle G.},
        title = "{The Athena++ Adaptive Mesh Refinement Framework: Design and Magnetohydrodynamic Solvers}",
      journal = {\apjs},
         year = 2020,
        month = jul,
       volume = {249},
       number = {1},
          eid = {4},
        pages = {4},
          doi = {10.3847/1538-4365/ab929b},
archivePrefix = {arXiv},
       eprint = {2005.06651},
 primaryClass = {astro-ph.IM},
       adsurl = {https://ui.adsabs.harvard.edu/abs/2020ApJS..249....4S}
}


\end{document}